\documentclass[twocolumn,floats,floatfix,superscriptaddress,aps,pra]{revtex4-1}
\usepackage{amsfonts,amssymb,amsmath}
\usepackage{calc,xcolor,esvect}
\usepackage{graphicx}
\usepackage{siunitx}
\usepackage{bm}
\usepackage[normalem]{ulem}

\def\be{ \begin{equation} }
\def\ee{ \end{equation} }
\def\bea{ \begin{eqnarray} }
\def\eea{ \end{eqnarray} }
\def\bse{ \begin{subequations} }
\def\ese{ \end{subequations} }
\def\sech{\,\text{sech}\,}

\def\i{\,\text{i}}
\def\e{\,\text{e}}
\def\i{i}
\def\e{e}

\def\to{\rightarrow}

\def\d{\text{d}}
\def\sech{\text{sech}}
\def\U{\mathbf{U}}

\def\to{\rightarrow}

\def\i{\text{i}}

\begin{document}

\author{Genko T. Genov}
\email{genko.genov@physik.tu-darmstadt.de}
\affiliation{Institut f{\"u}r Angewandte Physik, Technische Universit{\"a}t Darmstadt, Hochschulstr. 6, 64289 Darmstadt, Germany}
\author{Daniel Schraft}
\affiliation{Institut f{\"u}r Angewandte Physik, Technische Universit{\"a}t Darmstadt, Hochschulstr. 6, 64289 Darmstadt, Germany}
\author{Thomas Halfmann}
\homepage{http://www.iap.tu-darmstadt.de/nlq}
\affiliation{Institut f{\"u}r Angewandte Physik, Technische Universit{\"a}t Darmstadt, Hochschulstr. 6, 64289 Darmstadt, Germany}

\title{Rephasing efficiency of sequences of phased pulses in spin-echo and light-storage experiments}

\date{26 December 2018}

\begin{abstract}
We investigate the rephasing efficiency of sequences of phased pulses for spin echoes
and light storage by electro-magnetically induced transparency (EIT). We derive a simple theoretical model and show that the rephasing efficiency is very sensitive to the phases of the imperfect rephasing pulses. The obtained efficiency differs substantially for spin echoes and EIT light storage,
which is due to the spatially retarded coherence phases after EIT light storage. Similar behavior is also expected for other light storage protocols with spatial retardation or for rephasing of collective quantum states with an unknown/undefined phase, e.g., as relevant in single photon storage. We confirm the predictions of our theoretical model by experiments in a Pr$^{3+}$:Y$_{2}$SiO$_{5}$ crystal.
\end{abstract}

\maketitle

\section{Introduction}

Sensing, processing, and communication of quantum information
in realistic media usually suffers from dephasing processes due to unwanted interactions with the environment. Therefore, rephasing techniques are inherent part of many experimental protocols, e.g., for light storage \cite{Lvovsky09NPhot,Fleischhauer05RMP,MarangosHalfmann09Optics,Heinze13PRL,Lovric13PRL,Zhong15Nature,Jobez15PRL,Mieth16PRA,Schraft16PRL}, quantum sensing \cite{DegenRMP2017}, or quantum information \cite{HaeffnerPR2008,PiltzScience2016}.

Pulse imperfections are often a major limitation to high fidelity rephasing \cite{RDD_review12Suter}, leading to the development and implementation of robust rephasing schemes \cite{RDD_review12Suter,Torosov11PRA,Torosov11PRL,Schraft13PRA,Casanova15PRA,Van-Damme17PRA,Genov2014PRL,Genov17PRL}. 
Furthermore, the rephasing efficiency often depends on the initial quantum state of the system, e.g., with the widely used Carr‐Purcell‐Meiboom‐Gill (CPMG) sequence \cite{CPMG_papers}. Pulse error compensation, e.g., with composite pulses, can sometimes work better for some initial states, but not for others \cite{Levitt84NMR}.
Thus, the performance of rephasing sequences might differ in experiments where the initial state is usually known, e.g., spin echo, and when this is not the case, e.g., for quantum memories. Proper characterization of rephasing efficiency in the two cases is thus important. 

The performance of rephasing sequences
is usually analyzed theoretically by their single qubit fidelity \cite{RDD_review12Suter}.
The characterization is typically performed for rectangular pulses, specific errors (e.g., amplitude variation), and with the assumption that the qubits' phases are well-defined with respect to the rephasing pulses, e.g., in spin echo experiments \cite{RDD_review12Suter}.
Related theoretical investigations on the rephasing efficiency of collective atomic states with a single excitation were proposed recently \cite{Cruzeiro16JMO}.
To the best our knowledge, however, there are no investigations on the rephasing efficiency in the case of EIT light storage,
where the initial phase can vary along the atomic ensemble or is unknown/undefined.
This is especially important, e.g., for EIT quantum memories in atomic ensembles, where rephasing is applied to prolong storage time \cite{Lvovsky09NPhot,Heinze13PRL}.
There is also no explicit comparison with the rephasing efficiency in spin echo experiments.

In our work, we theoretically analyze the rephasing efficiency for spin echoes or EIT light storage. We confirm the theoretical findings by experiments in a doped solid.
First, we develop a simplified theoretical model for the rephasing efficiency of a sequence of imperfect rephasing pulses for (a) spin echoes or (b) EIT light storage. 
We then use it to derive explicit formulas for the rephasing efficiency of several example sequences.
The model specifies the performance in terms of the population transfer efficiency
and is, in principle, applicable to pulses of arbitrary shape.
It shows a substantial difference between the two cases, which can be explained by variation of the initial phases of atomic coherences after EIT light storage, e.g., due to spatial retardation. Such variation is also expected in other light storage protocols, as well as in the absence of retardation effects when the phases of qubits vary or are undefined, e.g., as in single photon storage.
Second, we performed an experimental investigation of rephasing efficiencies for spin echoes and EIT light storage in a Pr$^{3+}$:Y$_{2}$SiO$_{5}$ crystal (termed Pr:YSO hereafter).
We used several variants of phased ``detection'' sequences to demonstrate the differences between the two cases.
The experiments confirm the theoretical predictions, i.e., the rephasing efficiency is very sensitive to the phases of the imperfect rephasing pulses and differs significantly for spin echoes and EIT light storage.

Finally, we discuss an example of rephasing by CPMG, applied in EIT light storage. We show explicitly that CPMG with pulse errors cannot efficiently preserve an arbitrary initial quantum state and, hence, is not appropriate to rephase quantum memories.

\section{Theory}
%

\begin{figure*}
\includegraphics[width=0.85\textwidth]{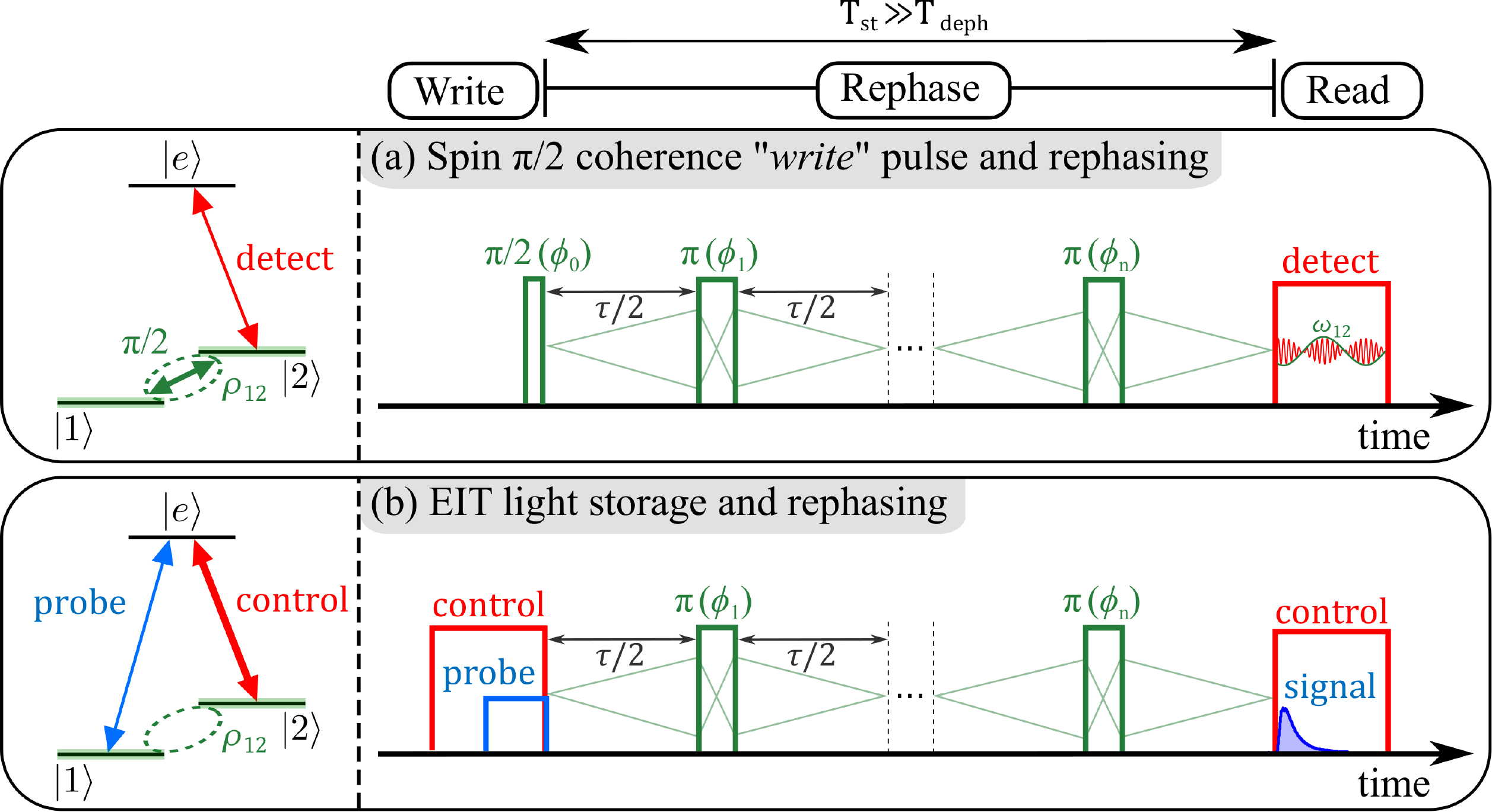}
\caption{(Color online)
Level schemes and schematic description of the rephasing experiments for (a) spin echoes or (b) EIT light storage. The system is initially prepared in state $|1\rangle$. Then we apply a ``write'' process, which creates atomic coherences by (a) a $\pi/2$ pulse on the $|1\rangle\leftrightarrow |2\rangle$ spin transition and (b) light storage of a weak probe field by electro-magnetically induced transparency (EIT). During the storage time $T_{\text{st}}$ the atomic coherences dephase due to inhomogeneous broadening of the spin transition. We perform rephasing by ideally resonant $\pi$ pulses to preserve the coherence for much longer than the dephasing time $T_{\text{deph}}$. We then measure (read) the experimental efficiency of the rephasing process by (a) a Raman heterodyne (RH) signal driven with a weak detection field or (b) a signal pulse driven by the control ``read'' pulse in EIT configuration.
}
\label{Fig:fig1}
\end{figure*}
\subsection{The System}
%
We consider an ensemble of non-interacting three-state systems, e.g., in a $\Lambda$-type atomic medium (see Fig. \ref{Fig:fig1}).
The quantum states $|1\rangle$ and $|2\rangle$ are assumed long-lived and thus suitable for optical data storage.
The individual atoms in the ensemble exhibit slightly different transition frequencies $\omega_{12}=\overline{\omega_{12}}+\Delta$, e.g., due to inhomogeneous broadening, where $\overline{\omega_{12}}$ is the center frequency of the ensemble and $\Delta$ is the frequency detuning of an individual atom.
All atoms are initially prepared in state $|1\rangle$ by an appropriate preparation, e.g., optical pumping.
Then, a \emph{``write''} process is applied,
and the atom states can be characterized by the density matrix $\rho(z,t=0)$,
where $z$ is the position of the atom and $t=0$ is the time immediately after the \emph{``write''} process. The parameters $\rho_{11}(z,t)$ and $\rho_{22}(z,t)$ show the populations in state $|1\rangle$ and $|2\rangle$, respectively, while $\rho_{12}(z,t)\equiv|\rho_{12}(z,t)|\exp{[\i\xi(z,t)]}$ is the slowly varying coherence of an atom in a frame rotating at an angular frequency $\overline{\omega_{12}}$, which has a phase $\xi(z,t)$.
For example, if the system is initially in state $|1\rangle$ and we apply a \emph{``write''} process by a resonant $\pi/2$ pulse with a phase $\phi_0$, see Fig. \ref{Fig:fig1}(a), the coherence phase afterwards is $\xi(z,t=0)=\phi_0+\pi/2$.
In a typical storage experiment, we are interested in the expectation value of the dipole moment, i.e., $\mu_{21}\langle\rho_{12}(z,t)\rangle$ + c.c. with
\begin{equation} \label{rho_def}
\langle\rho_{12}(z,t)\rangle= \int_{-\infty}^{+\infty}\rho_{12}(z,t) g(\Delta) \d \Delta,
\end{equation}
where $\mu_{21}=\langle 2|\mathbf{\mu}|1\rangle$ is the dipole moment on the $|1\rangle\leftrightarrow |2\rangle$ spin transition, 
$g(\Delta)$ describes the spectral distribution of the detunings in the inhomogeneous manifold of frequency ensembles, with $\int_{-\infty}^{+\infty} g(\Delta) \d \Delta=1$. We assume that $g(\Delta)$ has no spatial dependence and does not change with time. Then, if the system evolves freely after the \emph{``write''} process, we obtain
\begin{equation} \label{exp_rho_z}
\langle\rho_{12}(z,t)\rangle=\rho_{12}(z,0)\langle\e^{\i{\Delta t}}\rangle,
\end{equation}
with $\langle\e^{\i{\Delta t}}\rangle=\int_{-\infty}^{+\infty}\e^{\i{\Delta t}}g(\Delta) \d \Delta$.
It is obvious that the latter expression approaches $0$ for times much greater than a characteristic dephasing time $T_{\text{deph}}$ (depending on the spectral distribution $g(\Delta)$) unless rephasing pulses are applied.

\subsection{Rephasing efficiency model}

In order to counter the effect of dephasing, we apply \emph{``rephasing''} pulse(s) that enable preservation of $\langle\rho_{12}(z,t)\rangle$ for a storage time $T_{\text{st}}\gg T_{\text{deph}}$ (see Fig. \ref{Fig:fig1}, right).
Ideally, we use resonant pulse(s) on the $|1\rangle\leftrightarrow |2\rangle$ spin transition for rephasing. In the following, the notation $A(\phi)$ denotes a pulse with a target pulse area of $A$ and a relative phase $\phi$.
If (each of) the pulse(s) is resonant and has a pulse area of $\pi$ the effect of dephasing is reversed for every atom.
However, perfect resonant $\pi$-pulses are not possible in systems with large inhomogeneous broadening due to the different detuning $\Delta$ for the individual atoms.
The efficiency can be further reduced by spatial inhomogeneity of the applied field.
Hence, the driving pulse is no more a $\pi$-pulse for all atoms.
In order to investigate the rephasing efficiency for sequences of time-separated phased pulses,
we derive now a simplified theoretical model.

The density matrix after a rephasing process is
\begin{equation}\label{densityM_rephasing}
\rho(z,T_{\text{st}})=\mathbf{U}_{\text{reph}}(z,T_{\text{st}})\rho(z,0)\mathbf{U}_{\text{reph}}^{\dagger}(z,T_{\text{st}}),
\end{equation}
where $\mathbf{U}_{\text{reph}}$ is a propagator that depends on the applied rephasing sequence and can vary for each atom due to variation in the individual detuning $\Delta$ and/or the inhomogeneity of the field (see Appendix, Sec. \ref{Appendix:U_derivation}).

\begin{figure*}
\includegraphics[width=\textwidth]{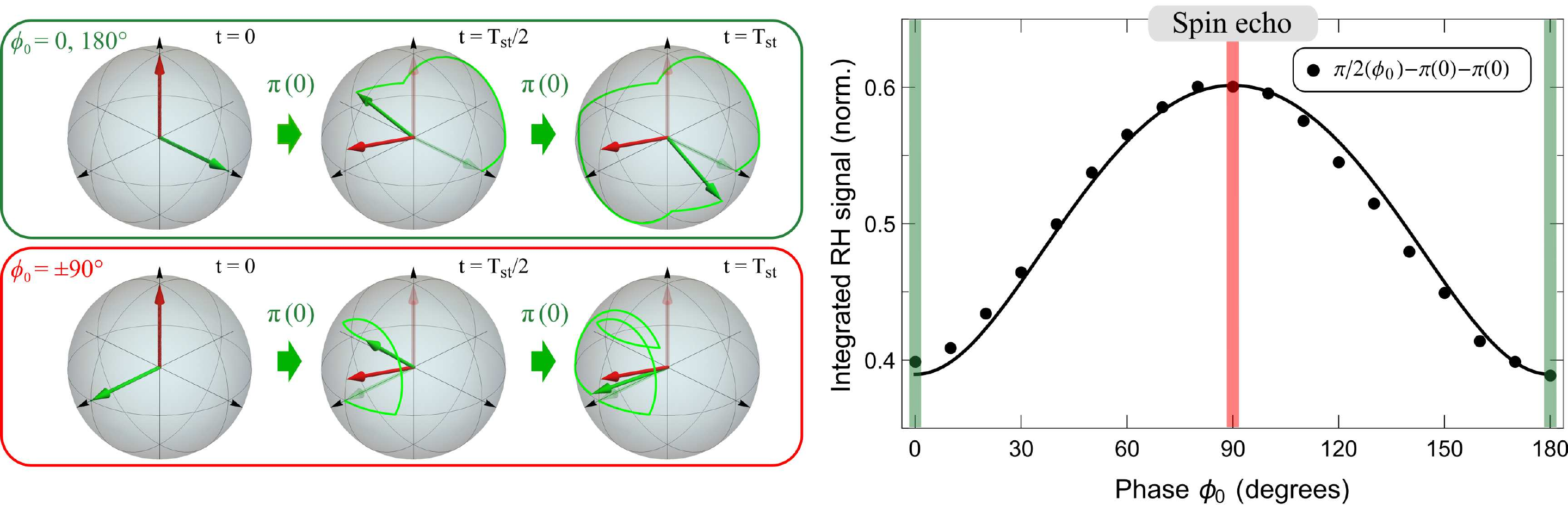}
\caption{(Color online)
(left) Bloch sphere representation of coherent evolution of the quantum state of an atom during a CPMG rephasing sequence of two time-separated pulses after a \emph{``write''} step with a perfect $\pi/2(\phi_0)$ pulse. The green (red) vector shows the Bloch (normalized torque) vector during the rephasing pulses. Rectangular pulses with a detuning $\Delta=0.2\Omega$, $\Omega=0.8\Omega$, $T_{\text{pulse}}=\pi/\Omega$, $\tau\approx 1.9 T_{\text{pulse}}$ were assumed with (left, top) $\phi_0=0, 180^{\circ}$, (left, bottom) $\phi_0=\pm 90^{\circ}$.
The error is smaller (and the efficiency higher) for $\phi_0=\pm 90^{\circ}$ when the error due to the first pulse is partially compensated by the second pulse.
(right) Experimentally measured rephasing efficiency vs. the phase $\phi_0$ of the $\pi/2$ \emph{``write''} pulse (black dots) and simulation (black line) for spin echoes in an atomic ensemble. The simulated rephasing efficiency is based on Eq. \eqref{Eq:CPMG_efficiency_RF}, $\epsilon=0.1$, normalized to the minimal and maximal values of the experimental data.
}
\label{Fig:fig2}
\end{figure*}

If we apply a single rephasing pulse with a relative phase $\phi_1$, i.e., a Hahn echo \cite{Hahn50PR,NMR_literature}, the coherence at the end of the process is given by Eq. \eqref{densityM_rephasing} and takes the form
\begin{subequations}
\begin{align}
\rho_{12}^{\text{Hahn}}&(z,T_{\text{st}})=a_0 +a_1 \exp{(\i\delta)}+a_2 \exp{(2\i\delta)},\\
a_0&=-(1-\epsilon)\e^{2\i(\beta+\phi_1)}\rho^{\ast}_{12}(z,0),\\
a_1&=\sqrt{\epsilon(1-\epsilon)}\e^{\i(\beta+\phi_1)}\left[\rho_{22}(z,0)-\rho_{11}(z,0)\right],\\
a_2&=\epsilon \rho_{12}(z,0),
\end{align}
\end{subequations}
where $\delta\equiv\alpha+\Delta \tau/2$ is an accumulated phase for the particular atom during the pulse and the free evolution times $\tau/2$ before and after the pulse.  The parameter $\epsilon\equiv 1-p$ is the error in the transition probability of the pulse for a particular atom with $p$ the probability that the atom is in state $|2\rangle$ after applying the pulse if it was initially in state $|1\rangle$. Finally, $\alpha$, $\beta$ are phases that depend on the pulse characteristics (see Appendix, Sec. \ref{Appendix:U_derivation}). We are interested in
\begin{subequations}\label{Eq:efficiency_approximation}
\begin{align}
\langle\rho^{\text{Hahn}}_{12}(z,T_{\text{st}})\rangle &\approx
\langle a_0 \rangle+\langle a_1\rangle \langle \exp{(\i\delta)}\rangle+\langle a_2 \rangle \langle\exp{(2\i\delta)} \rangle \\
\approx \langle a_0 \rangle&=-e^{\i\phi_1}\langle (1-\epsilon)\e^{2\i\beta} \rangle \rho^{\ast}_{12}(z,0),
\end{align}
\end{subequations}
where we used the approximation $\langle a_1\exp{(\i\delta)}\rangle\approx\langle a_1\rangle \langle \exp{(\i\delta)}\rangle$, i.e., $\exp{(\i\delta)}$ varies much faster than $a_1$ and $a_2$ when we perform the integration, defined in Eq. \eqref{rho_def}. This is feasible, e.g.,
if the time $\tau>T_{\text{deph}}$, which we use in the last step to make $\langle \exp{(\i\delta)}\rangle\to 0$. We note that the latter assumption might not hold when the time between the rephasing pulses is short, but even then the contribution of the higher order terms will be small when $\epsilon\to 0$. We also assumed for simplicity that the coherence $\langle\rho_{12}(z,0)\rangle\approx \rho_{12}(z,0)$.
We note the latter might not always be fulfilled, e.g., as the ``write'' pulse for spin echo is typically also imperfect. Nevertheless, we neglect this effect as it is usually small for the ratio $\langle\rho_{12}(z,t)/\rho_{12}(z,0)\rangle$, which we use in our analysis. Furthermore, we are interested in the efficiency of the rephasing pulses only.
Finally, we assumed that the phase $\phi_1$ is the same for all atoms, which is usually feasible.

We derive expressions for the coherences after rephasing for sequences of $n$ time-separated rephasing pulses in a similar way
\begin{equation}
\langle\rho_{12}(z,T_{\text{st}})\rangle =
\langle a_0 \rangle+\sum_{m=1}^{2n}\langle a_{m}\exp{(\i m\delta)}\rangle\approx \langle a_0 \rangle,
\end{equation}
neglecting all higher order terms of the expansion $\langle a_{m}\exp{(\i m\delta)}\rangle \to 0$ due to the fast variation in $m\delta$ for the different atoms at position $z$ in the medium in comparison to $a_{m}$. This is our main assumption that allows significant simplification. It is usually feasible for time separation $\tau>T_{\text{deph}}$ or $\epsilon\to 0$.

In the next subsections, we use our theoretical approach to investigate the rephasing efficiency in two different cases: (a) for atomic coherences with phases well defined with respect to the subsequent imperfect rephasing $\pi$ pulses, e.g., as in spin echo experiments (see Fig. \ref{Fig:fig1}a), and (b) light storage, e.g., by electro-magnetically induced transparency (EIT), where the relative phase of the created coherence is usually varying/unknown with respect to the phases of the rephasing pulses (see Fig. \ref{Fig:fig1}b).

\subsection{Rephasing efficiency for spin echo experiments}

We assume now, that the coherence phase $\xi(z,0)$ after a \emph{``write''} process is the same for all atoms in the storage medium and it is well defined with respect to the phases of subsequent rephasing pulses. For example, if the system is initially in state $|1\rangle$ and we apply a $\pi/2(\phi_0)$ pulse on the $|1\rangle\leftrightarrow|2\rangle$ transition, with a pulse duration much shorter then the dephasing time, and pulse wavelength much longer than the medium length, we have $\xi(z,0)\approx\phi_0+\pi/2$ all across the medium.
We determine the rephasing efficiency by the ratio
\be\label{Eq:Efficiency_RF}
\widetilde{\eta}_{\text{r}}
=\left\vert\frac{\langle\rho_{12}(z,T_{\text{st}})\rangle}{\rho_{12}(z,0)}\right\vert,
\ee
This definition is an appropriate choice also for our experimental implementation, which uses Raman-heterodyne (RH) detection. The magnitude of the detected RH amplitude is proportional to the magnitude of the coherence at the end of the storage time \cite{Wong83PRB}.

The efficiency of rephasing with a single pulse (usually termed Hahn echo) is obtained from \eqref{Eq:efficiency_approximation} and takes the form 
\be\label{Eq:Hahn_efficiency_RF}
\widetilde{\eta}_{\text{r}}^{\text{Hahn}}=\left\vert\left\langle (1-\epsilon\right)\e^{2\i\beta}\rangle\right\vert,
\ee
where the brackets imply averaging of the respective parameter for the different atoms in the ensemble, e.g., $\langle \epsilon \rangle = \int \epsilon g(\Delta)\d \Delta$ is the averaged error in the transition probability because of variation in the detuning $\Delta$, e.g., due to inhomogeneous broadening. We note, that averaging over variations of the Rabi frequency or other experimental parameters can also be taken into account in a similar way.
Then, a rephasing pulse that performs perfect population transfer $\epsilon=0$ for all atoms would yield $\widetilde{\eta}_{\text{r}}^{\text{Hahn}}=1$, assuming small variation in $\beta$. This is the case for rectangular rephasing pulses, i.e., $\beta=-\pi/2$ (see Appendix, Sec. \ref{Appendix:U_derivation}), which we assume further on for simplicity.
We note for completeness, that the assumption of small variation of $\beta$ is not always feasible, e.g., when we apply a chirped rephasing pulse. Then, $\beta$ can vary significantly because of the dynamic phase due to the chirped pulse (see Appendix, Sec. \ref{Appendix:U_derivation}). Thus, the efficiency of Hahn echo suffers, which has also been confirmed in previous experiments \cite{Lauro11PRA,Mieth12PRA,Pascual-Winter13NJP}.

We apply the same approach to calculate the rephasing efficiency for the widely used CPMG sequence \cite{CPMG_papers}. It usually consists of two time-separated resonant pulses, each ideally with a pulse area of $\pi$, which we denote $[\pi(\phi_1)-\pi(\phi_2)]$.
We note that rephasing with CPMG can also be applied with adiabatic chirped pulses, where each performs population inversion \cite{Schraft13PRA}. Then, the pulse area is usually much greater than $\pi$ to satisfy the adiabatic condition \cite{Schraft13PRA,Mieth12PRA,Shore_literature}.
We also note that throughout this work we refer to spin echo as an experiment where there is a well-defined relation between the phase of the first coherence creation $\pi/2$ pulse and the phases of the subsequent rephasing $\pi$-pulses. The spin echo signal can be obtained with different rephasing $\pi$-pulse sequences, e.g., CPMG with two $\pi$-pulses (shown in Fig. \ref{Fig:fig2}), and does not refer solely to the Hahn sequence where we use a single rephasing $\pi$-pulse.
We choose $\phi_1=0$ without loss of generality and obtain 
\begin{equation}\label{Eq:CPMG_efficiency_RF}
\widetilde{\eta}_{\text{r}}^{\text{CPMG}}=\left\vert\left\langle (1-\epsilon)\left[1+\epsilon-2\epsilon\left(1+\e^{-\i(2\phi_0+\phi_2)}\right)\right]\right\rangle\right\vert.
\end{equation}
Again, a rephasing pulse that performs perfect population transfer ($\epsilon=0$) yields $\widetilde{\eta}_{\text{r}}^{\text{CPMG}}=1$. Unlike Hahn echo, this condition is also sufficient for chirped pulses where the dynamic phase from the first pulse is canceled by the second pulse \cite{Lauro11PRA,Mieth12PRA,Pascual-Winter13NJP}.

There is also another difference compared to the Hahn echo: the rephasing efficiency with imperfect pulses now depends on the phase relation between the \emph{``write''} pulse $\phi_0$ and the phase $\phi_2$ of the second rephasing $\pi$-pulse. When $2\phi_0+\phi_2=0$, the efficiency is lowest and given by
\begin{equation}\label{CPMG_efficiency_bad}
\widetilde{\eta}_{\text{r}}^{\text{CPMG}}=\left\vert\left\langle (1-\epsilon)(1-3\epsilon)\right\rangle\right\vert.
\end{equation}
The original Carr-Purcell \cite{CPMG_papers} sequence $\pi/2(0)-\pi(0)-\pi(0)$ is an example for this case. We note that a phase $\phi_k=0$ of a pulse ideally implies rotation around the the $X$ axis of a Bloch sphere (see Fig. \ref{Fig:fig2} (left)). Thus, if the system was initially in state $|1\rangle$ and we apply a $\pi/2(\phi_0=0)$ \emph{``write''} pulse, the Bloch vector will then point along the $Y$ axis of the Bloch sphere and the ``bad'' initial coherence phase will be $\xi(z,0)\approx \phi_0+\pi/2=\pi/2$. Figure \ref{Fig:fig2} (left, top)) provides intuition for the underlying reason for the worse performance for the Carr-Purcell sequence.
Then, the pulse error of the second rephasing $\pi(0)$ pulse adds up to the error of the first $\pi(0)$ pulse.

When the $2\phi_0+\phi_2=\pi$, the efficiency is improved and given by
\begin{equation}\label{CPMG_efficiency_good}
\widetilde{\eta}_{\text{r}}^{\text{CPMG}}=\left\langle 1-\epsilon^2\right\rangle.
\end{equation}
The improved CPMG sequence, originally proposed by Meiboom and Gill \cite{CPMG_papers}, $\pi/2(\pi/2)-\pi(0)-\pi(0)$, is an example of this case. 
After a $\pi/2(\phi_0=\pm\pi/2)$ \emph{``write''} pulse, the Bloch vector will point along the X axis of the Bloch sphere and the ``good'' initial coherence phase will be $\xi(z,0)\approx \phi_0+\pi/2=0$ or $\pi$. Figure \ref{Fig:fig2} (left, bottom)) provides intuition for the improved performance. The quantum state after the \emph{``write''} $\pi/2$ pulse is closely aligned with the torgue vector of the rephasing pulses, so the second rephasing $\pi(0)$ pulse partially compensates the error of the first $\pi(0)$ pulse for the particular initial state along the $X$ axis.

Our analysis so far showed that CPMG efficiency depends on the value of $2\phi_0+\phi_2$. This result implies that choosing a different phase of the second rephasing $\pi$-pulse cannot improve performance for an arbitrary initial coherence phase. A different choice of $\phi_2$ only shifts the ``good'' value of $\phi_0$ (and thus of $\xi(z,0)$).
For example, one can naively think that the CPMG-2 sequence $\pi(0)-\pi(\phi_2=\pi)$, also termed $X-(-X)$, would perform better than CPMG 
as the second $\pi$-pulse of CPMG-2 has ideally an opposite rotation axis to the first $\pi$-pulse.
Indeed, CPMG-2 can compensate pulse area errors for arbitrary states but only for resonant pulses and negligible dephasing between the pulses.
However, its error compensating mechanism fails in the case of significant dephasing during or between the imperfect $\pi$-pulses, e.g., due to inhomogeneous broadening.
In the latter case, which we analyze, the sequence $\pi/2(0)-\pi(0)-\pi(\pi)$ has improved efficiency as $2\phi_0+\phi_2=\pi$ and the second rephasing $\pi(\pi)$ pulse partially compensates the error of the first $\pi(0)$ pulse. In other words, CPMG-2 has improved performance when the initial Bloch vector after the \emph{``write''} pulse points along the Y axis of the Bloch sphere, so the ``good'' initial coherence phase is $\xi(z,0)=\pm \pi/2$. However, the sequence $\pi/2(\pi/2)-\pi(0)-\pi(\pi)$ now has a worse efficiency as $2\phi_0+\phi_2=0$ (mod $2\pi$). Thus, we achive a worse performance with CPMG-2 when the initial Bloch vector points along the X axis of the Bloch sphere, so the ``bad'' initial coherence phase is now $\xi(z,0)=0$ or $\pi$. As the ``good'' and ``bad'' initial coherence phases are only shifted in comparison to standard CPMG, rephasing by CPMG-2 will have the same efficiency when averaged over different $\phi_0$ (and thus over $\xi(z,0)$).

In summary, our analysis showed that $\pi(0)-\pi(\phi_2)$ has improved performance when $2\phi_0+\phi_2=\pi$ (mod $2\pi$), i.e., the ``good'' initial coherence phase is $\xi(z,0)=\phi_2/2+\pi k,k \in \mathbb{Z}$. However, its performance suffers when $2\phi_0+\phi_2=0$ (mod $2\pi$), i.e., for the ``bad'' initial coherence phase $\xi(z,0)=(\pi+\phi_2)/2+\pi k,k \in \mathbb{Z}$.

The performance of CPMG for $\phi_0=0$ and different $\phi_2$ is demonstrated further below in the experimental section and confirms our theoretical predictions. 
We can derive analytical formulas also for three and more pulses, but will not discuss the rather complicated features here.
The simulated rephasing patterns for a sequence of four pulses is also shown further below
in the experimental section.

Finally, we analyze the case for repeated application of CPMG, e.g., for dynamical decoupling.
We denote the CPMG sequence, repeated $N$ times, as $[\pi(0)-\pi(0)]^N$. The repetition of CPMG with imperfect pulses will lead to even more pronounced differences in the rephasing efficiencies vs. phase of the initial coherence, as shown in Eqs.\eqref{CPMG_efficiency_bad} and \eqref{CPMG_efficiency_good}. Assuming a small variation in $\epsilon$ ($\epsilon\ne 0$) for the different atoms and $N\to\infty$, it is possible to show that
\begin{subequations}\label{Efficiency_many_CPMG}
\begin{align}
\widetilde{\eta}_{\text{r}}&\to 0,~~~~~~~~~~~~~~\xi(z,0)=\pi/2+\pi k,\\
\widetilde{\eta}_{\text{r}}&\approx 1-\langle\epsilon\rangle/\sqrt{2},~~\xi(z,0)=\pi k,
\end{align}
\end{subequations}
where in the last approximation we neglected the effect of the higher moments of $\epsilon$, which is usually valid, e.g., when $\langle\epsilon^k\rangle\ll\langle\epsilon\rangle,~k=2,3,\dots$ and $\epsilon\lessapprox 0.5$. Thus, rephasing of the ``good'' phase does not suffer from repeated application of the CPMG sequence, also with pulse errors. Every second pulse approximately cancels the error of the previous pulse for the specific initial state, determined by the rotation axis of the CPMG pulses (see Fig. \ref{Fig:fig2}(left, bottom)) \cite{DegenRMP2017,RDD_review12Suter,NMR_literature}. Thus, applying many CPMG sequences with pulse errors in an ensemble effectively projects (spin-locks) the quantum state of the atoms onto this quantum state, i.e., onto the state with the ``good'' phase.

\begin{figure}[t]
\includegraphics[width=\columnwidth]{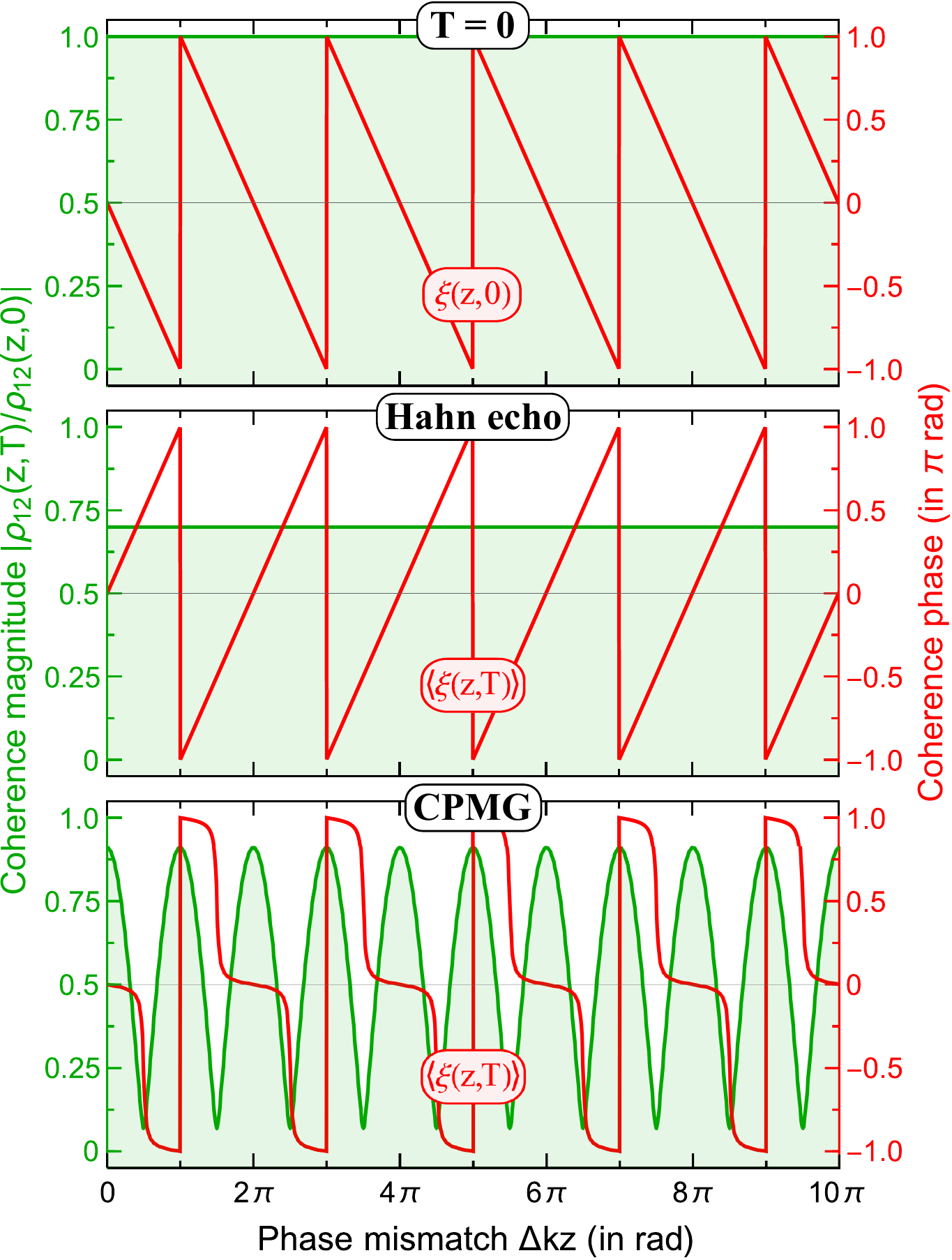}
\caption{(Color online)
Theoretical simulations for the coherence magnitude (light green) and phase (red) vs. the phase mismatch in EIT light storage: (top) after the \emph{``write''} step, (middle) after rephasing with a single Hahn echo pulse, (bottom) after rephasing with a CPMG sequence of two pulses. The coherence phase $\xi(z,0)=-\Delta k z$ is assumed to vary significantly along the propagation axis of the probe field with $\Delta k L=10\pi$, where $L$ is the length of the storage medium. The shading in light green at different $z$ is added to guide the eye. The theoretical simulations are performed for $\epsilon=0.3$, based on Eq. \eqref{Eq:efficiency_approximation}. We note the inverted phase for a Hahn echo, and the grating pattern in the coherence magnitude for CPMG of two pulses $\pi(0)-\pi(0)$, as its rephasing efficiency changes due to the variation of the initial coherence phase.
}
\label{Fig:fig3}
\end{figure}

\subsection{Rephasing efficiency for EIT light storage}\label{Subsec:Theory_LS_Efficiency}

Control over the phase of the initial coherences after the \emph{``write''} process cannot be achieved with light storage where the phase $\xi(z,0)$ is unknown with respect to the applied rephasing pulses and/or is spatially retarded. This is typical for many light storage protocols, e.g., atomic-frequency combs (AFC) or electromagnetically-induced transparency (EIT). Our analysis focusses on EIT light storage, i.e., when the coherence phase $\xi(z,0)$ varies due to spatial retardation. We will discuss other possible scenarios in section \ref{Sec:Discussion}.

\begin{figure}[t]
\includegraphics[width=\columnwidth]{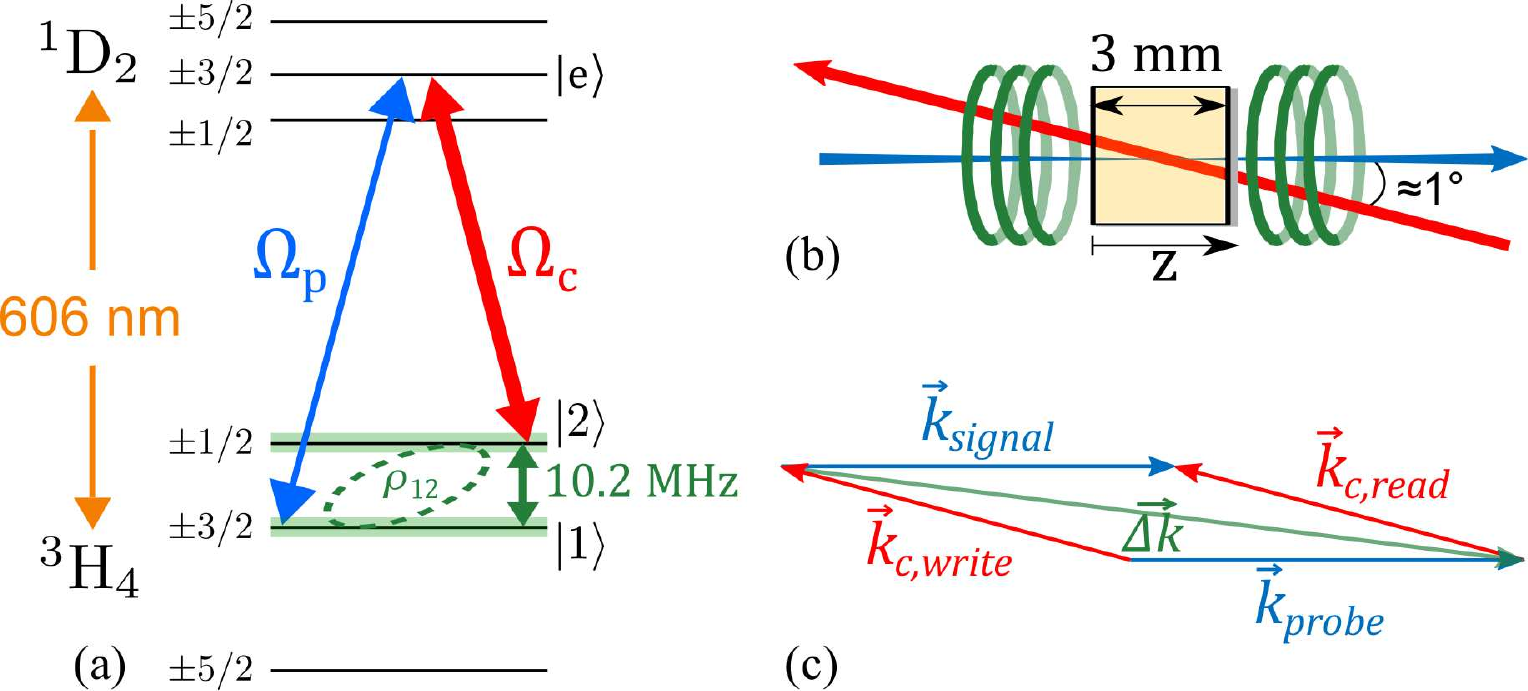}
\caption{(Color online)
(a) Level scheme of Pr$^{3+}$ ions, as relevant for our experimental implementation in a Pr:YSO crystal. A spin coherence is prepared on the $|1\rangle\leftrightarrow |2\rangle$ transition by a $\pi/2$ rf pulse or EIT light storage. (b) Simplified geometry of the optical beam paths and rf coils around the Pr:YSO crystal in the experiment.
The angle between the probe and control beams is approximately 1$^\circ$, which ensures sufficient overlap in the crystal. However, the magnitude of the wave vector
$\protect\vv{\Delta k}$
is large as the two beams propagate in opposite directions.
(c) Respective phase-matching condition for EIT light storage and retrieval.
}
\label{Fig:fig4}
\end{figure}

Light storage by EIT is implemented in a $\Lambda$-type atomic medium (see Fig. \ref{Fig:fig1}(b)), where the atoms are initially prepared with all population in the ground state $|1\rangle$ \cite{Fleischhauer05RMP,MarangosHalfmann09Optics,Novikova12review}. EIT uses a strong control pulse, tuned to the transition between the ground state $|2\rangle$ and an excited state $|e\rangle$.
The probe and control pulse are assumed phase coherent, e.g., they can be derived from the same laser source.
The coherent interaction with the control pulse makes the medium transparent for a probe pulse on the $|1\rangle \leftrightarrow|e\rangle$ transition and reduces its group velocity. Thus, the probe pulse is compressed in the storage medium.
By reducing the control pulse intensity adiabatically, the ``slow light'' probe pulse is ``stopped'' and converted into an atomic coherence of the quantum states $|1\rangle$ and $|2\rangle$ along the probe propagation path $z$ in the atomic medium. This establishes a spin wave of spatially distributed atomic coherences $\rho_{12}(z,0)$ in the medium, which contains all information of the incoming probe pulse. This is usually termed the \emph{``write''} process of EIT light storage and in the perfect case it maps \cite{Fleischhauer05RMP}
\begin{equation}
\mathcal{E}_{\text{probe}}(z,0)\to\sqrt{N/V}\rho_{12}(z,0)\e^{\i\Delta k z},
\end{equation}
where $\mathcal{E}_{\text{probe}}(z,t)$ is the electric field envelope of the probe pulse, $N/V$ is the number density of atoms, $\rho_{12}(z,t)$ is the coherence at position $z$ in the ensemble, $\Delta k=(\mathbf{k}_{\text{probe}}-\mathbf{k}_{\text{c,write}})_{z}$ is the phase mismatch between probe and control \emph{``write''} beams along the $z$ propagation axis of the probe beam, e.g., due to the geometry of the experiment (see also Fig. \ref{Fig:fig4}(b,c)).
We assume without loss of generality that $\mathcal{E}_{\text{probe}}(z,0)$ is real.
The coherence phase $\xi(z,0)=-\Delta k z$ is spatially retarded and vary along $z$ (see also Fig. \ref{Fig:fig3}, top).

During a storage time $T_{\text{st}}$, we \emph{``rephase''} our coherences by applying a sequence of time-separated pulses analogously to the previous section.
The spin wave is then read out by applying a control read pulse to beat with the atomic coherences and generate a signal pulse on the $|1\rangle\leftrightarrow|e\rangle$ transition.
\begin{equation}
-\sqrt{N/V}\langle\rho_{12}(z,T_{\text{st}})\rangle\e^{\i\Delta k z}\to \mathcal{E}_{\text{signal}}(z,T_{\text{st}}).
\end{equation}

The light storage efficiency is typically determined by the ratio of the energy of retrieved photons after storage vs. the energy of the photons of the input probe pulse \cite{Fleischhauer05RMP,Novikova12review}.
\begin{equation}
\eta_{\text{ls}}=\frac{\int_{T_{\text{st}}}^{\infty}| \mathcal{E}_{\text{signal}}(z=L,t)|^2\d t}{\int_{-\infty}^{0}|\mathcal{E}_{\text{probe}}(z=0,t)|^2\d t},
\end{equation}
where $z=0$ and $z=L$ are the beginning and end of the storage medium.
If the mapping of the probe field to atomic coherences during the \emph{``write''} and \emph{``read''} steps is perfect, the overall light storage efficiency would correspond to the rephasing efficiency and we obtain for the latter (see Appendix, Sec. \ref{Appendix:EIT_reph_derivation})
\begin{equation}\label{LS_efficiency_all}
\eta_{\text{ls}}=\frac{|\int_{0}^{L}\langle\rho_{12}(z,T_{\text{st}})\rangle\e^{\i\Delta k z} \d z|^2}{|\int_{0}^{L}\rho_{12}(z,0)\e^{\i\Delta k z}\d z|^2}.
\end{equation}
We note, that in case of imperfect pulses the overall light storage efficiency would be lower than the rephasing efficiency, e.g., due to reabsorption of the probe pulse as the system will not be perfectly aligned with the dark state, required for EIT, during readout.

We can further simplify Eq. \eqref{LS_efficiency_all} by change of variables $\Delta k z\to -\xi(z,0)$ and by assuming that the phases of the coherence after the \emph{``write''} process $\xi(z,0)$ are equally distributed between $0$ and $2\pi$. The latter is a valid assumption in the limit of large spatial retardation $\Delta k L\gg 2\pi$ (see Fig. \ref{Fig:fig3}), which is often feasible, e. g., due to
the geometry of the experimental setup when there is an angle between the probe and control fields. We also assume for simplicity that $|\rho_{12}(z,0)|$ does not vary much along the usually small distance where $\Delta k z$ changes from $0$ to $2\pi$. Thus, the rephasing efficiency in the approximation of equal coherence phase distribution becomes (see Appendix, Sec. \ref{Appendix:EIT_reph_derivation})
\begin{equation}\label{LS_efficiency_simple}
\eta_{\text{ls}}=\left|\frac{1}{2\pi}\int_{0}^{2\pi}\frac{\langle\rho_{12}(z,T_{\text{st}})\rangle}{\rho_{12}(z,0)} \d \xi(z,0)\right|^2.
\end{equation}
Next, we use the above analytical formulas to estimate the efficiency for several imperfect rephasing sequences applied for EIT light storage.

First, we analyze the rephasing efficiency of a single rephasing pulse, i.e., the well-known Hahn echo \cite{Hahn50PR} for EIT light storage. Figure \ref{Fig:fig3} shows a simulation for the spatial variation of the magnitude of the coherence ratio $\langle\rho_{12}(z,T_{\text{st}})\rangle/\rho_{12}(z,0)$ and the phase $\langle\xi(z,T_{\text{st}})\rangle$ along the propagation axis of the probe pulse after Hahn echo rephasing. As noted in the previous section, the magnitude of the final coherence after Hahn echo does not depend on the initial phase. However, the final phase is inverted, i.e., $\langle\xi(z,T_{\text{st}})\rangle=-\xi(z,0)$. Then, phase matching implies that the individual atomic dipoles along the axis of probe field propagation will not be properly phased during standard forward readout and, thus, their emission cannot add up coherently in the propagation direction of the signal field. We then use Eqs. \eqref{Eq:efficiency_approximation} and \eqref{LS_efficiency_all} to estimate the rephasing efficiency
\begin{align}\label{LS_efficiency_Hahn}
\eta^{\text{Hahn}}_{\text{ls}}&=\left|\frac{1}{L}\int_{0}^{L}\left\langle(1-\epsilon)\right\rangle\e^{2\i\Delta k z} \d z\right|^2
\notag\\&
\approx\left\langle(1-\epsilon)\right\rangle^2\left|\frac{\e^{2\i\Delta k L}-1}{2\Delta k L}\right|^2,
\end{align}
where again $\epsilon\equiv 1-p$ is the error in the transition probability of the rephasing pulse for a single atom and we assumed that $\epsilon$ varies much more slowly that $\Delta k z$ in the last calculation step. We also assumed that $\beta=-\pi/2$ for any $z$. This is the case when we apply a short rectangular-shaped pulse in time, and the frequency of the $|1\rangle\leftrightarrow |2\rangle$ transition is in the radio frequency (rf) regime, i.e., the wavelength is much longer than the medium. In the limit of large spatial retardation $\Delta k L\gg 2\pi$ the rephasing efficiency becomes
\begin{equation}
\eta^{\text{Hahn}}_{\text{ls}}\to 0.
\end{equation}
We note that rephasing by a Hahn echo is also expected to work inefficiently for light storage because of reabsorption of the signal field during retrieval, as the populations of states $|1\rangle$ and $|2\rangle$ are interchanged and the system will not be in the dark state, required for EIT.

Next, we estimate the rephasing efficiency for the CPMG sequence $[\pi(0)-\pi(\phi_2)]$. Figure \ref{Fig:fig3} shows an example for the spatial variation of coherences along the propagation axis of the probe pulse after CPMG with pulse errors ($\phi_2=0$), simulated from Eq. \eqref{Eq:efficiency_approximation}. The rephasing efficiency varies along $z$ as the performance of CPMG is then highly dependent on the initial coherence phase. As a result, the final coherence magnitude exhibits a distinct grating-like pattern, with the highest magnitude corresponding to the ``good'' phases $\langle\xi(z,T_{\text{st}})\rangle\to 0$ or $\pi$. Additionally, the phases of the final coherences also tend to these ``good'' phases. Thus, applying a CPMG sequence ($\phi_2=0$) with pulse errors effectively partially projects the quantum states of the individual atoms onto the quantum state, determined by the rotation axis of the CPMG pulses, making $\langle\xi(z,T_{\text{st}})\rangle\to 0$ or $\pi$. We use Eqs. \eqref{Eq:efficiency_approximation} and \eqref{LS_efficiency_simple}
to obtain the overall rephasing efficiency for light storage
\begin{equation}\label{LS_efficiency_CPMG}
[\pi(0)-\pi(\phi_2)]:~\eta^{\text{CPMG}}_{\text{ls}}=\left\langle(1-\epsilon)^2\right\rangle^2.
\end{equation}
It is notable, that the light storage rephasing efficiency does not depend on the phase $\phi_2$ of the second rephasing pulse - in contrast to the spin echo case (see Eqs. \eqref{CPMG_efficiency_bad} and \eqref{CPMG_efficiency_good}). For example, the performances of CPMG ($\phi_2=0$) and CPMG-2 ($\phi_2=\pi$) rephasing sequences applied to light storage are equal while they worked differently for spin echoes. The reason is that the rephasing efficiency for EIT light storage is averaged over all possible coherence phases in the medium, e.g., due to the spatial retardation.

Similarly, we can estimate the rephasing efficiencies for sequences of three and more time-separated pulses. We apply the same assumptions as for CPMG and obtain
\begin{align}
&[\pi(0)-\pi(\phi_2)-\pi(0)]:~\eta_{\text{ls}}=\left\langle 4\epsilon(1-\epsilon)^2\cos{(\phi_2)}\right\rangle^2,\\
&[\pi(0)-\pi(\phi_2)]^2:\\ &\eta_{\text{ls}}=\left\vert\left\langle(1-\epsilon)^2\left(e^{4\i\phi_2}(1-\epsilon)^2
+6\epsilon^2+4\e^{2\i\phi_2}\epsilon(2\epsilon-1)\right)\right\rangle\right\vert^2,\notag
\end{align}
where the index in $[\pi(0)-\pi(\phi_2)]^2$ denotes twice application of the CPMG sequence. The rephasing efficiency of sequences of more pulses and with other phases can be calculated in an analogous way.
Explicit analytical formulas for several robust sequences are included in Table \ref{Table:sequences} in the Appendix, Sec. \ref{Appendix:Efficiency_robust_sequences}.

Finally, we consider the rephasing efficiency of CPMG, repeated many times, e.g., for dynamical decoupling. The repeated sequence CPMG $[\pi(0)-\pi(0)]^N$ leads to more pronounced differences in the rephasing efficiency vs. initial coherence phase, as shown in Eqs.\eqref{CPMG_efficiency_bad} and \eqref{CPMG_efficiency_good}. Assuming a small variation in $\epsilon$ for the different atoms,
we deduce the rephasing efficiency for $[\pi(0)-\pi(0)]^N$, applied to light storage, for very large $N$ as
\begin{equation}
[\pi(0)-\pi(0)]^N:~\eta_{\text{ls}}\approx\left\langle \frac{1-\epsilon/\sqrt{2}}{2}\right\rangle^2
\end{equation}
In the above approximation we neglected the effect of higher moments of $\epsilon$, similarly to the spin echo case. Thus, for large $N$, the number of repetitions of the CPMG sequence with pulse errors does not matter. The often repeated CPMG sequence effectively projects (spin-locks) the quantum states of the atoms onto the quantum state, determined by the rotation axis of the CPMG pulses, i.e., onto a state with the ``good'' phase (see Eqs. \eqref{Efficiency_many_CPMG}).

In summary, we derived a simplified theoretical model for the rephasing efficiency of sequences of pulses in atomic ensembles in two cases:
(a) when the individual atoms' quantum states exhibit the same, well-defined phase with respect to the rephasing pulses, e.g.,
in a spin echo experiment,
and (b) when the phases of the individual coherences vary, e.g., due to spatial retardation, for light storage by electro-magnetically induced transparency (EIT). We find, that the rephasing efficiency is very sensitive to the phases of the imperfect rephasing pulses. The behaviour of the rephasing efficiency differs significantly for spin echoes or EIT light storage.
In the following, we verify the theoretical predictions by rephasing experiments for spin echo and EIT light storage in a doped solid.

\begin{figure*}
\includegraphics[width=0.99\textwidth]{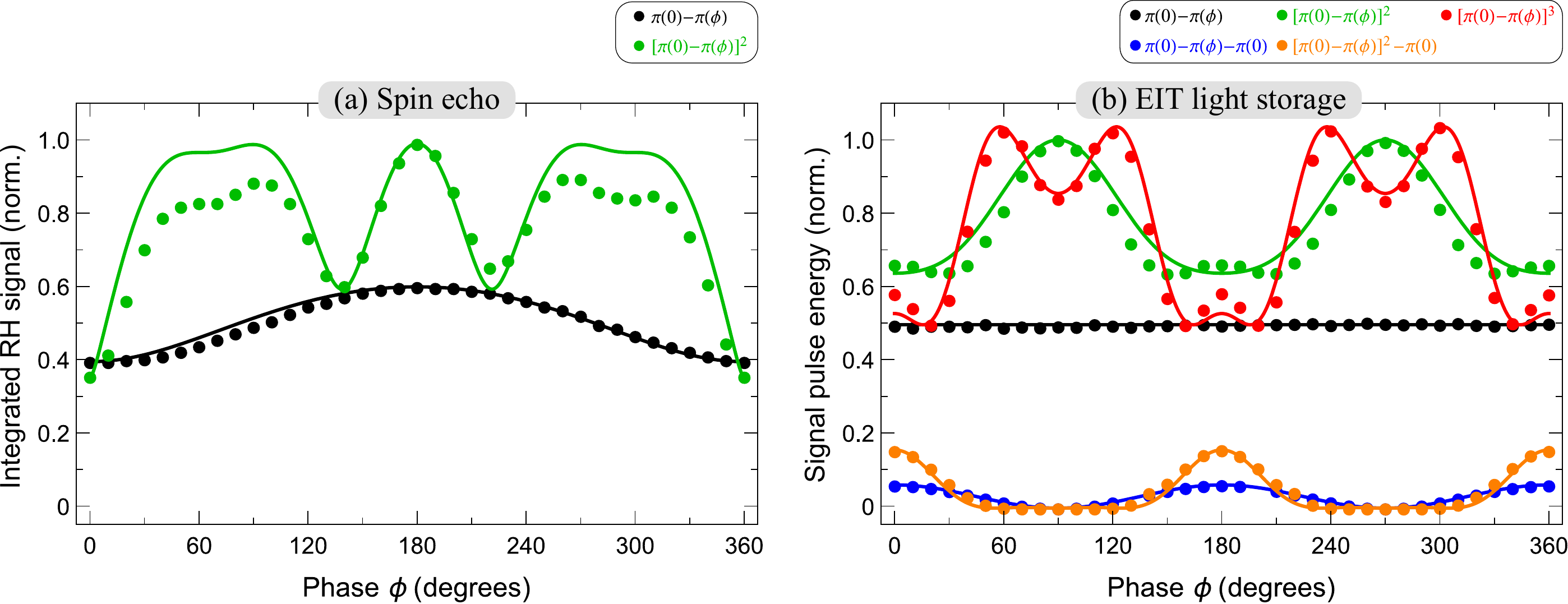}
\caption{(Color online)
Experimentally measured rephasing efficiency (dots) and theoretical simulations (lines) vs. the phase $\phi$ of every second rephasing pulse
for (a) spin echo after a $\pi/2(\phi_0=0)$ \emph{``write''} pulse
and (b) EIT light storage. We apply different rephasing sequences (indicated by different colours in the graphs). 
The simulations use $\epsilon=0.1$ and are normalized to the extrema of the experimental data for each rephasing sequence. The experimental data are calibrated to the extrema of the $[\pi(0)-\pi(\phi)]^2$ rephasing sequence. 
}
\label{Fig:fig5}
\end{figure*}

\section{Experimental demonstration}
The experiments are performed in a Pr:YSO crystal with a length of 3.2 mm and a dopant concentration of $0.05$ at.$\%$ praseodymium. The crystal is mounted in a liquid-helium cryostat and held at temperatures of about $4$ K to reduce phononic excitations. Figure \ref{Fig:fig4}(a) shows the relevant part of the level scheme of a Pr$^{3+}$ ion. The optical transition between the electronic ground state $^{3} H_{4}$ and the excited state $^{1} D_{2}$ is at a wavelength of 605.98 nm.
The population lifetime of the ground state $^{3} H_{4}$ is of the order of $T_{1}\approx 100$ s \cite{Heinze13PRL,Heinze14PRA}.
Local variations in the crystal field for the different Pr$^{3+}$ ions lead to inhomogeneous broadening of both the optical ($\Gamma_{\text{opt}}\approx 7$ GHz) and spin transitions.
The latter is of the order of kHz and leads to a dephasing time of the spin coherence on the hyperfine transitions of the order of
$T_{\text{deph}}\equiv T_{2}^{*}\approx 10~\mu$s.
Additionally, stochastic magnetic interactions between the dopant ions and the host matrix lead to a decoherence time of $T_{2}\approx 500~\mu$s \cite{Schraft13PRA,Mieth16PRA}. 
Figure \ref{Fig:fig4}(b) shows a simplified setup of the optical beam path and rf coils next to the crystal in the experiment. For more details on the experimental setup, including the generation of optical and rf pulses, as well as the optical preparation of the medium see \cite{Schraft13PRA,Mieth14OE}.

Each experimental cycle starts with an optical pump sequence to prepare the system in the initial state $|1\rangle$. Afterwards, we perform three steps, i.e., \emph{``write''}, \emph{``rephasing''} and \emph{``read''} (see Fig. \ref{Fig:fig1}). The \emph{``write''} and \emph{``read''} processes are different for (a) spin echo and (b) EIT light storage, while the \emph{``rephasing''} processes are the same for both storage protocols.

In the spin echo rephasing experiments, we apply a radio frequency (rf) $\pi/2$ pulse with a phase $\phi_0$ during the \emph{``write''} step to generate a maximum coherent superposition between states $\vert 1\rangle$ and $\vert 2 \rangle$ (see Fig. \ref{Fig:fig1}(a)).
Unless explicitly otherwise noted, we store the atomic coherences for a storage time of $T_{\text{storage}}=600~\mu$s, which is much longer than the dephasing time of $T_{\text{deph}}\approx 10~\mu$s.
In order to reverse the effect of dephasing during the storage time, in the \emph{``rephasing''} step we apply ideally resonant rf pulses, each with a pulse area of $\pi$. The relative phases between the pulses serve as control parameters to drive and compare different types of rephasing sequences, as discussed in the theory section above.
Since we ``wrote'' the atomic coherence by a rf pulse with full experimental control of the waveform, we have a precisely defined relative phase between the atomic coherences after the \emph{``write''} step and the rf rephasing pulses \cite{Schraft13PRA,Heinze13PRL}. We note, that such precisely defined relative phases between storage and rephasing pulses are, in principle, also possible in appropriate setups with optical pulses \cite{Serrano2018NattComm}.
Unless otherwise noted, the rephasing pulses in all measurements have a rectangular shape in time, a target pulse area of $\pi$, pulse duration of $T=3.2\mu$s, corresponding to a Rabi frequency of $\Omega\approx 2\pi 156$ kHz, optimized for the highest light storage efficiency with CPMG. The time separation between the pulses is variable, depending on the specific sequence, while the latter always fit in the fixed storage time $T_{\text{st}}$. In the \emph{``read''} step for spin echoes, we use an optical detection field
to detect and measure the spin coherence on the $\vert 1\rangle\leftrightarrow\vert 2 \rangle$ by RH detection \cite{Wong83PRB,Heinze14PRA}.
The detection field scatters at the spin coherence to generate a Stokes field on the $\vert 1\rangle$ and $\vert e \rangle$ transition. Stokes and detection field interfere with each other to provide a beating pattern, which we observe on a photo diode and demodulate by a lock-in amplifier. The magnitude of the signal is proportional to the magnitude of the coherence on the $|1\rangle\leftrightarrow |2\rangle$ transition at the end of the storage time \cite{Wong83PRB}. We note, that the rephasing efficiencies are reduced due to inhomogeneous broadening of the spin transition and
the spatial inhomogeneity of the rf field along the crystal.

For EIT light storage, we \emph{``write''} atomic coherences on the $\vert 1\rangle\leftrightarrow\vert 2 \rangle$ transition by Raman-type two-photon interaction of a weak classical probe pulse and a strong classical control pulse in the $\Lambda$-type level scheme (see Fig. \ref{Fig:fig1}(b)).
Figure \ref{Fig:fig4}(c) shows the phase matching condition for EIT light storage and retrieval. The geometry of the experiment leads to a large $\Delta k=(\mathbf{k}_{\text{probe}}-\mathbf{k}_{\text{c,write}})_{z}$, i.e., phase mismatch between to probe and control \emph{``write''} beams along the $z$ propagation axis of the probe beam.
The coherence phase $\xi(z,0)$ after EIT light storage then varies in the storage medium due to spatial retardation along the propagation axis of the probe pulse. The effect is quite large in our specific experiment due to the counter-propagating probe and control beams,
i. e., $\Delta k\approx 2k_{\text{probe}}$, so $\Delta k L\approx 10^7$, where $L\approx 3.2$ mm is the crystal length.
(see Fig. \ref{Fig:fig4}(b,c)).
The \emph{``rephasing''} sequences for EIT light storage are the same as in the case of spin echoes. In order to \emph{``read''} the optical memory, we again apply the strong optical control \emph{``read''} pulse in the same direction as in the \emph{``write''} step
(forward readout). Despite the large $\Delta k$, if the readout phase matching condition $\mathbf{k}_{\text{signal}}=\mathbf{\Delta k}+\mathbf{k}_{\text{c,write}}$ is satisfied, the control \emph{``read''} pulse beats with the atomic coherences and generates a signal pulse in the same direction as the probe field (see also Fig. \ref{Fig:fig4}(c)) \cite{Surmacz08PRA,Schraft16PRL}.

We present now measurements of the rephasing efficiency for different rephasing sequences and compare them with the theoretical results, in both cases of spin echoes and EIT light storage. In the first experiment we investigate spin echoes, applying a rf $\pi/2$ \emph{``write''} pulse of a pulse duration $1.6\mu$s to drive maximal atomic coherences. We varied the phase $\phi_0$ of the rf \emph{``write''} pulse, while keeping the phases of all subsequent rephasing pulse(s) constant (we take $\phi_{k}=0,k\ne 0$ without loss of generality).  The experimentally obtained rephasing efficiency vs. phase of the initial coherence is shown in Fig. \ref{Fig:fig2}, along with the theoretically expected dependence. As expected,
the rephasing efficiency of CPMG varies significantly with $\phi_0$. It reaches a maximum for $2\phi_0+\phi_2=\pi$, i.e., $\phi_0=\pm \pi/2$ and a minimum for $2\phi_0+\phi_2=2\pi k$, i.e., $\phi_0=\pi k, k\in \mathbb{Z}$. The experimental data fit very well the simulation based on Eq. \eqref{Eq:CPMG_efficiency_RF} with $\langle\epsilon\rangle=\epsilon=0.1$ (for simplicity we assumed $\langle\epsilon^k\rangle\approx\langle\epsilon\rangle^k$).
We note, that our simulations are normalized to the extrema of the experimental data to exclude perturbing effects beyond dephasing, e.g., stochastic phase fluctuations. In Pr:YSO, the latter is caused by random changes in the transition frequencies due to spin flips in the host lattice \cite{Lvovsky09NPhot,Heinze13PRL}.

In a modified version of the spin echo experiment, we again applied a rf $\pi/2$ \emph{``write''} pulse, but kept its phase $\phi_0=\phi_{2k+1}=0,k\in \mathbb{N}$. Instead, we varied now the phases $\phi_{2k}\equiv\phi$ of all even $\pi$ pulses in the rephasing sequences. Fig. \ref{Fig:fig5}(a) shows the experimental results for single and double application of a GPMG sequence $[\pi(0)-\pi(\phi_2)]$, i.e., with two or four pulses. Similarly to the previous experiment, the rephasing efficiency of CPMG varies clearly with phase $\phi_2$. For the single two-pulse CPMG sequence, we observe a maximum for $2\phi_0+\phi_2=\pi$, i.e., $\phi_2=\pm \pi$, and a minimum for $2\phi_0+\phi_2=2\pi k$, i.e., $\phi_2=2\pi k, k\in \mathbb{Z}$ (see black data points in Fig. \ref{Fig:fig5}(a)).
The dependence of the rephasing efficiency vs. phase becomes more complicated for a double four-pulse CPMG sequence, showing several and very pronounced extrema (see green data points in Fig. \ref{Fig:fig5}(a)). As expected, repeated application of CPMG sequences yields a much stronger variation of the performance with phase. Also here, the simulations confirm the experimental findings. We note, that the rephasing efficiency for a double CPMG sequence is higher compared to a single CPMG sequence - although pulse errors are expected to add up for the longer sequence. However, the improvement is due to the shorter time separation between the pulses for the longer sequence in order to keep the total duration of any sequence in the fixed storage time. The shorter pulse separation reduces the effect of stochastic phase fluctuations between the pulses in a sequence \cite{Genov17PRL}.

We shift our attention now to rephasing in EIT light storage experiments. Similarly to the spin echo measurement discussed before, we varied the phases $\phi_{2k}\equiv\phi$ of all even pulses in CPMG $[\pi(0)-\pi(\phi_2)]$ rephasing sequences, while keeping $\phi_{2k+1}=0$ fixed. Figure \ref{Fig:fig5}(b) shows the results of these measurements. We first consider application of the single two-pulse CPMG sequence (see black data points in Fig. \ref{Fig:fig5}(b)). The dependence upon the phase is flat, i.e., very different to spin echoes. As already discussed in theory section, this difference is due to averaging over all possible coherence phases in the medium generated by EIT light storage involving spatial retardation. However, the flat dependence changes for the double, i.e., four-pulse, CPMG sequence (see green data points in Fig. \ref{Fig:fig5}(b)). It exhibits a strong variation of the rephasing efficiency with phase - though still different from spin echoes. Also a triple, i.e., six-pulse, CPMG sequence (see red data points in Fig. \ref{Fig:fig5}(b)) reveals similar oscillation of the rephasing efficiency.
The data fit well to the numerical simulations of all sequences, i.e., confirming our theoretical model.

\begin{figure}[t]
\includegraphics[width=0.95\columnwidth]{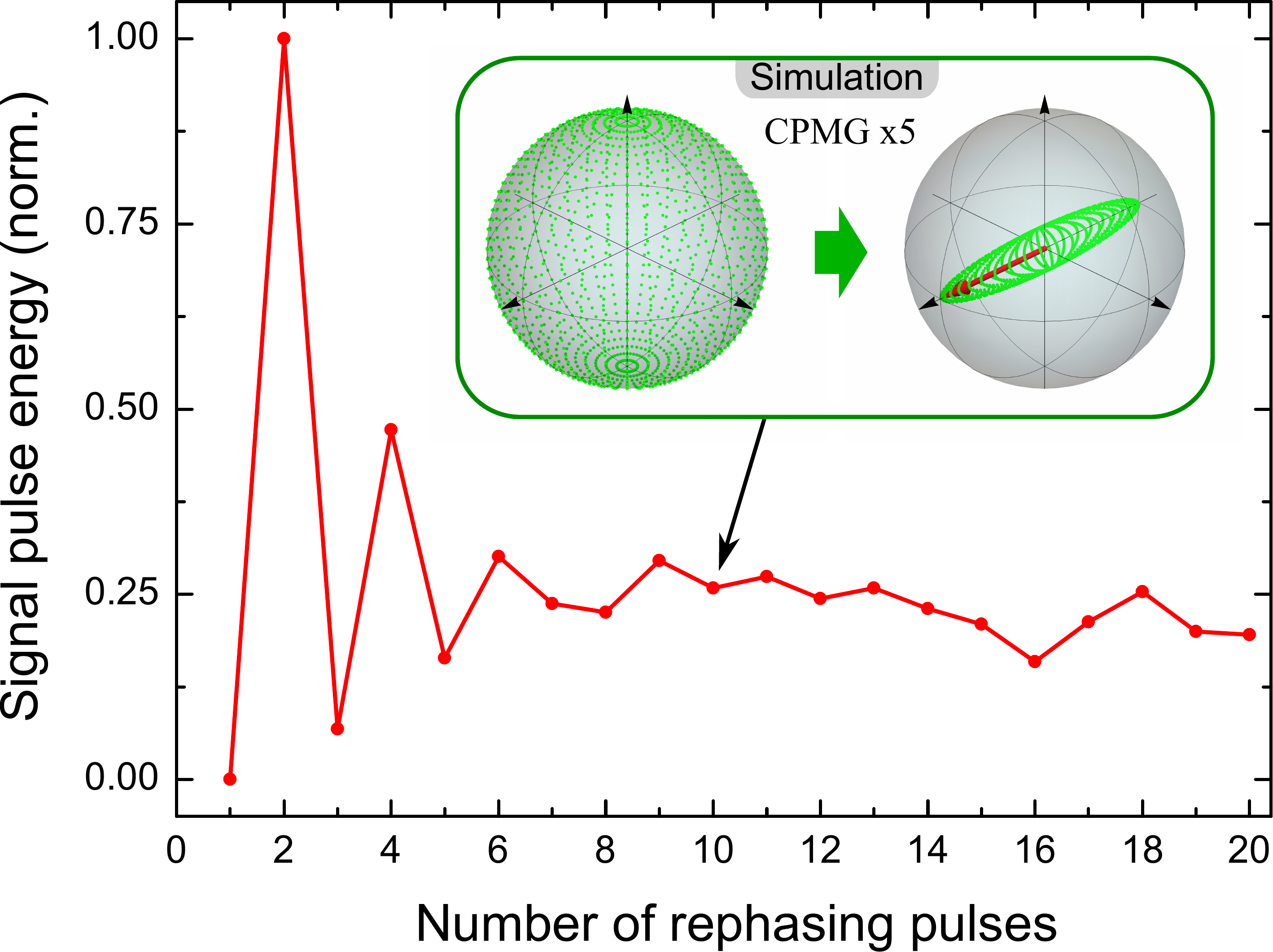}
\caption{(Color online)
Experimentally measured signal pulse energy vs. number of rephasing pulses for a total storage time of $T_{\text{st}}=300~\mu$s. The data are calibrated with respect to the signal with CPMG with two rephasing pulses. The rephasing pulses exhibit rectangular time profile with a duration of $T=3.3~\mu$s and a Rabi frequency of $\Omega\approx 2\pi~156$ kHz. All pulses have the same relative phase, assumed to be zero without loss of generality.
\emph{Inset:} Simulation of rephasing on the Bloch sphere for an ensemble of atoms, initially in the same state, for different initial states, depicted by green dots (see Bloch sphere on the left). When we apply a CPMG sequence 5 times, i.e., with a total of 10 pulses, the averaged Bloch vector for the ensemble is projected along the red arrow, which depicts the average (for the ensemble), normalized torque vector during a pulse (see Bloch sphere on the right). The simulation assumes pulses with the same properties as in the experiment (see above), and a cycle time of $T+\tau=30\mu$s, so that we fit the 10 pulses within the storage time of $T_{\text{st}}=300 \mu$s. Detuning variation with a bandwidth of $\approx \pm 2\pi~40$ kHz is assumed.
}
\label{Fig:fig6}
\end{figure}

%
In a concluding measurement we varied the number of rephasing pulses with zero phases for a total storage time of $T_{\text{st}}=300~\mu$s.
According to theory, sequences with an odd number of perfect $\pi$ pulses (e.g., a single pulse as in a Hahn echo) yield no light storage signal at all. This is due to phase-mismatch and inversion of the ground states populations, as already discussed in the theory section. Hence, only sequences with an even number of pulses are applicable for quantum memory in our experiment.
Figure \ref{Fig:fig6} confirms our theoretical predictions that we indeed obtain no signal with a single rephasing pulse.
However, we obtain a small signal for sequences with a higher number of odd pulses, e.g., three, five, which is due to pulse errors.
We also found, that the light storage efficiency for repeated CPMG sequences, i.e., of more than 10 single pulses, does not depend significantly on whether we use an odd or an even number of them. This confirms the prediction that CPMG is significantly affected by pulse errors.
After several repetitions the quantum state of each atom is ``projected'' (spin-locked). This is confirmed by a numerical simulation (see the inset of Fig. \ref{Fig:fig6}), which shows the average Bloch vector of an atomic ensemble (green dot) for different initial states on the Bloch sphere. As noted in the theory section, the overall light storage efficiency is expected lower than the maximal rephasing efficiency in Eq. \eqref{LS_efficiency_CPMG}, due to reabsorption of the retrieved signal, which is not taken into account in our model. After the ``projection'', the rephasing efficiency of CPMG at the specific storage time does not depend on the number of pulses. We note that a similar spin-lock effect exists with the CPMG-2 but the projection is on the $Y$ axis of the Bloch sphere.

\section{Discussion}\label{Sec:Discussion}
In our analysis, we assumed that the phase of the atomic coherence $\xi(z,0)$ varies substantially in the medium, due to spatial retardation along the propagation axis of the probe pulse. The latter effect is quite large in our specific experiment, due to the counter-propagating probe and control beams.
In case of co-propagating beams, the angle between the two beams can realistically be of the order of at least $\theta\approx 1^{\circ}$ (see Fig. \ref{Fig:fig4}(b)), which implies $\Delta k L\approx 5>\pi$, where $\Delta k\approx k_{\text{pr}}(1-\cos{\theta})$, $L=3.2$ mm. Hence, in this case spatial retardation will be smaller, but still significant.
The above findings are also relevant to other light storage protocols involving retardation effects, e.g., atomic frequency combs \cite{Jobez15PRL}), by taking into account the specific characteristics of the protocol, e.g., phase matching conditions.

Moreover, our approach is applicable also beyond the specific effect of spatial retardation. As an example we note that the estimation of rephasing efficiencies by integration over all possible initial coherence phases $\xi(z,0)$, similarly to Eq. \eqref{LS_efficiency_simple}, is also applicable when the initial coherence phase is not defined. This is relevant for rephasing of collective quantum states with a well defined number of excitations \cite{PezzeRMP2018}, e.g., for highly symmetric Dicke states after single or few photon storage. Specifically, it is well known that an incoherent average over the phase $\xi(z,0)$ of a collective Bloch state can be used to represent a mixture of symmetric Dicke states \cite{Arecchi72PRA,Friedberg07LPL}. The statistical distribution of the latter can be sharply peaked when the number of atoms is large and the number of collective excitations is small. Thus, it is justified to approximate the rephasing efficiency for a collective entangled state after single photon storage with the rephasing efficiency of a phase-averaged Bloch state, as shown in \cite{Cruzeiro16JMO}. Our theoretical approach is applicable also in these cases.

\section{Conclusion}
In conclusion, we investigated the efficiency of imperfect rephasing pulse sequences in atomic ensembles
for (a) spin echoes and (b) light storage by electro-magnetically induced transparency (EIT) in a doped solid crystal. We developed a simplified theoretical model for the rephasing efficiency in both cases. We found, that the rephasing efficiency varies considerably with the relative phase between the atomic coherences and the applied rephasing pulses. Moreover, the behaviour of the rephasing efficiency vs. phase differs significantly for coherences generated by spin echoes or EIT light storage.
While spin echoes provide well-defined phases with respect to the rephasing pulses, in EIT light storage spatial retardation of the phases of individual coherences plays a crucial role. We confirmed the theoretical predictions by experimental implementations of rephasing for spin echoes and EIT light storage in a Pr:YSO crystal. The data clearly prove the differences between rephasing for spin echoes and EIT light storage. We compared the behaviour of sequences of CPMG pulse pairs, either with pulses of equal phases or phase shifts inbetween, as well as longer pulse sequences with a variable number of pulse pairs. Finally, we demonstrated the spin-locking effect of dynamical decoupling with CPMG pulse pairs in EIT light storage. CPMG dynamical decoupling with pulse errors permits long storage times of selected coherences, but is not appropriate to rephase states with an arbitrary phase, as relevant, e.g., in quantum memories. The experimental data fit well with numerical simulations based on our theoretical model. The findings are of relevance also for other light storage protocols, whenever the initial phase of the quantum state varies along the storage medium, or whenever the initial phase is not precisely defined.

\begin{acknowledgments}
The authors thank B. W. Shore, N. V. Vitanov, M. Hain and S. Mieth for useful discussions. This work is supported by the Alexander von Humboldt Foundation and a Career Bridging Grant of the Technische Universit{\"a}t Darmstadt.
\end{acknowledgments}

\section{Appendix}
%

\subsection{Derivation of the propagator for a rephasing sequence}
\label{Appendix:U_derivation}

We depict rephasing with a single pulse, i.e., the well-known Hahn echo \cite{Hahn50PR}, by the schematic representation $[\tau/2-A(\phi)-\tau/2]$,
where the label $A(\phi)$ indicates a pulse with area $A$ and phase $\phi$, and $\tau/2$ is the duration of free evolution before and after the pulse. We depict longer sequences, e.g., a CPMG sequence with two time-separated pulses \cite{CPMG_papers}, by $[\tau/2-A(\phi_1)-\tau/2]-[\tau/2-A(\phi_2)-\tau/2]$. In the following, we compress this notation to $[A(\phi_1)-A(\phi_2)]$. Finally, the notation $[A(\phi_1)-A(\phi_2)]^N$ implies the sequence $[A(\phi_1)-A(\phi_2)]$ is repeated $N$ times.

The evolution due to the pulse $A(\phi=0)$ is described by a propagator $\U_{\text{pulse}}(0)$, which connects the atom density matrix at the initial and final times $\mathbf{\rho}(t_f)=\mathbf{U}(0) \mathbf{\rho}(t_i)\mathbf{U}^{\dagger}(0)$ (without loss of generality, we take $\phi=0$ as a global phase without physical relevance). The propagator $\U_{\text{pulse}}(0)$ can be conveniently parameterized with three real variables $p$ ($0\leqq p \leqq 1$), $\alpha$ and $\beta$ as
\begin{equation} \label{Upulse}
\U_{\text{pulse}}(0) = \left[\begin{array}{cc} \sqrt{1-p}\, \e^{\i\alpha}  & \sqrt{p}\, \e^{\i\beta} \\  -\sqrt{p}\, \e^{-\i\beta} & \sqrt{1-p}\, \e^{-\i\alpha} \end{array} \right],
\end{equation}
where the phases $\alpha$, $\beta$ depend on the pulse properties and $p$ is the transition probability induced by the pulse, i.e., the probability that the system is in state $|2\rangle$ after the pulse, when it was initially in state $|1\rangle$. We define the error in the transition probability $\epsilon\equiv 1-p$.
Table \ref{Table:pulse_propagator} shows examples for these variables for several conventional rephasing pulses, i.e., a resonant pulse, a detuned rectangular pulse and an adiabatic chirped pulse (assuming coherent evolution, dipole and rotating-wave approximations \cite{Shore_literature,Demkov-Kunike,GenovOptComm2011}).

Free evolution of an atom with a transition angular frequency $\overline{\omega_{12}}+\Delta$, where $\overline{\omega_{12}}$ is the center frequency of an ensemble of atoms and $\Delta$ is the frequency detuning of the individual atom, is described in the rotating frame at a frequency $\overline{\omega_{12}}$ by the propagator
\begin{equation}
\mathbf{F}(\Delta) = \left[\begin{array}{cc} \e^{\i\Delta\tau/4}  & 0, \\
0 & \e^{-\i\Delta\tau/4} \end{array} \right],
\label{eq:free_evol}
\end{equation}
where $\tau/2$ is the duration of the free evolution, e.g., before or after the rephasing pulse.

The propagator of the rephasing cycle $[\tau/2-A(\phi)-\tau/2]$ by a single time-separated, phase-shifted pulse for a particular atom then takes the form
\begin{align}
\mathbf{U}(\phi)&=\mathbf{F}(\Delta) \U_{\text{pulse}}(\phi) \mathbf{F}(\Delta) \notag\\
&= \left[\begin{array}{cc} \sqrt{1-p}\, \e^{\i\delta}  & \sqrt{p}\, \e^{\i(\beta+\phi)} \\  -\sqrt{p}\, \e^{-\i(\beta+\phi)} & \sqrt{1-p}\, \e^{-\i\delta} \end{array} \right],
\label{eq:CPMG-cycle}
\end{align}
where $\delta\equiv\alpha+\Delta \tau/2$ is an accumulated phase for the particular atom during the pulse and the free evolution times $\tau/2$ before and after the pulse. We note that the parameters $p$, $\alpha$, $\beta$ are affected, e.g., by detuning $\Delta$, field inhomogeneities, etc.
We also note that the propagator $\mathbf{U}(\phi)$, which takes free evolution into account,
has the same transition probability as the propagator of the rephasing pulse $\mathbf{U_{\text{pulse}}}(\phi)$ and differs only by the transformation $\alpha\to\delta$.
The propagator of a rephasing sequence of $n$ pulses with relative phases $\phi_1, \phi_2,\dots, \phi_{n}$ is
\be
\mathbf{U}_{\text{reph}}=\mathbf{U}(\phi_{n})\mathbf{U}(\phi_{n-1})\dots\mathbf{U}(\phi_{1}),
\ee
where we assumed that each rephasing cycle, including the pulse errors, can be repeated and we have control over the relative phase shifts between the pulses.

The density matrix of an atom after a storage time $T_{\text{st}}$ 
takes the form
\begin{equation}
\rho(z,T_{\text{st}})=\mathbf{U}_{\text{reph}}(z,T_{\text{st}})\rho(z,0)\mathbf{U}_{\text{reph}}^{\dagger}(z,T_{\text{st}}),
\end{equation}
where $\mathbf{U}_{\text{reph}}(z,T_{\text{st}})$ is a propagator that depends on the applied rephasing sequence and can vary for each atom due to variation in the individual detuning $\Delta$ and/or the inhomogeneity of the field, e.g., along $z$.

\begin{table}[t]
\caption{Elements of the pulse propagator in Eq.\eqref{Upulse} for resonant, rectangular and adiabatic chirped pulses \cite{Shore_literature,Demkov-Kunike,GenovOptComm2011}.
} 
\begin{tabular}{l} 
\hline 
\hline 
Resonant pulse, $\Omega(t)$: any shape, $t\in [0,T]$, $\Delta(t)=0$:\\ 
$p=1-\epsilon=\sin^2{(A/2)}$,\qquad\qquad\qquad\qquad~$A\equiv \int_{0}^{T}\Omega(t)\d t$\\
$\alpha=0$, $\beta=-\pi/2$\\
\hline 
Rectangular pulse, $\Omega(t)=\Omega,~t\in [0,T]$, $\Delta(t)=\Delta$:\\ 
\begin{math}
\begin{aligned}[t]
&p=1-\epsilon=\frac{\Omega^2}{\Omega^2+\Delta^2}\sin^2{(A_{\text{eff}}/2)},\qquad A_{\text{eff}}\equiv T\sqrt{\Omega^2+\Delta^2}\\
&\alpha=\tan^{-1}\left(\frac{\Delta}{\sqrt{\Omega^2+\Delta^2}}\tan{(A_{\text{eff}}/2)}\right),\qquad\alpha\in[-\pi/2,\pi/2]\\
&\beta=-\pi/2
\end{aligned}
\end{math}\\
\hline 
Adiabatic chirped pulse \cite{Demkov-Kunike,GenovOptComm2011},\\
$\Omega(t)=\Omega_0~\sech(t/T)$, $\Delta(t)=\Delta_0+B_0\tanh(t/T)$,\\
$t\in [-t_{\text{f}},t_{\text{f}}],~t_{\text{f}}\to\infty$:\\ 
\begin{math}
\begin{aligned}[t]
&U_{11}=\exp{(2\i\widetilde{D} t_{\text{f}}/T)}\frac{\Gamma(\nu)\Gamma(\nu-\lambda-\mu)}{\Gamma(\nu-\lambda)\Gamma(\nu-\mu)},~~~~\\
&U_{12}=\frac{-\i\widetilde{A} 2^{2\i\widetilde{B}}}{1-\nu}\exp{(2\i\widetilde{B} t_{\text{f}}/T)}\frac{\Gamma(2-\nu)\Gamma(\nu-\lambda-\mu)}{\Gamma(1-\lambda)\Gamma(1-\mu)},~~~~\\
&p=1-\epsilon=|U_{12}|^2,~\alpha=\arg{(U_{11})},~\beta=\arg{(U_{12})},\\
&\widetilde{A}\equiv\Omega_0 T/2,~\widetilde{B}\equiv B_0 T/2,~\widetilde{D}\equiv\Delta_0 T/2,~\\
&\lambda\equiv\sqrt{\widetilde{A}^2-\widetilde{B}^2}-\i\widetilde{B},~
\mu\equiv -\sqrt{\widetilde{A}^2-\widetilde{B}^2}-\i\widetilde{B},\\
&\nu\equiv \frac{1}{2}+\i(\widetilde{D}-\widetilde{B})
\end{aligned}
\end{math}\\
\hline 
\hline 
\end{tabular}
\label{Table:pulse_propagator} 
\end{table}

Then, the density matrix for an ensemble of atoms 
at position $z$ after a rephasing sequence of pulses is
\begin{equation}
\langle\rho(z,T_{\text{st}})\rangle=\int_{-\infty}^{+\infty}\rho(z,T_{\text{st}})g(\Delta)\d \Delta,
\end{equation}
where $g(\Delta)$ is the already defined spectral distribution of the detunings of the individual atoms.

A single pulse, which performs perfect population inversion ($\epsilon=0$) for every atom, e.g., a perfect resonant pulse $\pi(\phi_1)$, has a propagator
\be
\mathbf{U}_{\epsilon=0}(\phi)=\left[\begin{array}{cc} 0  & \e^{\i(\beta+\phi_1)} \\  -\e^{-\i(\beta+\phi_1)} & 0 \end{array} \right].
\ee
Then, the density matrix after a Hahn echo is
\be
\rho(z,T_{\text{st}})=\left[\begin{array}{cc} \rho_{22}(z,0)  & -\e^{2\i(\beta+\phi_1)}\rho_{21}(z,0) \\  \e^{-2\i(\beta+\phi_1)}\rho_{12}(z,0) & \rho_{11}(z,0) \end{array} \right],
\ee
i.e., the populations are inverted due to the single rephasing pulse, and the coherence $\rho_{12}(z,T_{\text{st}})=-\e^{2\i(\beta+\phi_1)}\rho_{12}(z,0)^{\ast}$.
Thus, the magnitude of the coherence at position $z$ is preserved, but its phase is inverted and shifted, i.e., $\langle\xi(z,T_{\text{st}})\rangle=\langle-\xi(z,0)+2\phi_1+2\beta+\pi\rangle$ (see also Fig. \ref{Fig:fig3}). We note that $\beta=-\pi/2$ for a resonant or rectangular pulse, which allow for spin echo rephasing with a single pulse. However, $\beta$ can vary significantly for chirped pulses because of their dynamic phase (see Table \ref{Table:pulse_propagator}). Thus, the efficiency of Hahn echo suffers, which has also been confirmed in previous experiments \cite{Lauro11PRA,Mieth12PRA,Pascual-Winter13NJP}.

This problem does not occur for CPMG with two time-separated rephasing pulses with relative phases $\phi_1$ and $\phi_2$, respectively, which perform perfect population inverstion ($\epsilon=0$). Then, the density matrix after rephasing is
\be
\rho(z,T_{\text{st}})=\left[\begin{array}{cc} \rho_{11}(z,0)  & \e^{-2\i(\phi_1-\phi_2)}\rho_{12}(z,0) \\  \e^{2\i(\phi_1-\phi_2)}\rho_{12}(z,0) & \rho_{22}(z,0) \end{array} \right],
\ee
i.e., it is identical to the initial matrix, except for a constant phase shift $-2(\phi_1-\phi_2)$ of the coherence, which is the same for every atom. We note that the density matrix does not depend on $\beta$, so rephasing with chirped pulses becomes possible, as shown in experiments \cite{Lauro11PRA,Mieth12PRA,Pascual-Winter13NJP}.

However, perfect rephasing pulses, e.g., resonant $\pi$ pulses, are not possible in systems with large inhomogeneous broadening due to the different detuning $\Delta$ for the individual atoms.
The efficiency may further decrease due to spatial inhomogeneity of the applied field. In these cases,
the driving pulse is no longer a $\pi$ pulse for all atoms. The rephasing efficiency depends then on the initial state of the atoms, e.g., on the coherence phase.

\subsection{Detailed derivation of the rephasing efficiency in EIT light storage}
\label{Appendix:EIT_reph_derivation}

Control over the phase of the initial coherences after the \emph{``write''} process cannot be achieved in EIT light storage. In this case, the phase $\xi(z,0)$ is unknown with respect to the phase of the applied rephasing pulses and/or is spatially retarded.
For example, light storage by EIT is usually implemented in a $\Lambda$-type atomic medium (see Fig. 1, right for an idealized scheme), where the atoms are initially prepared with all population in the ground state $|1\rangle$ \cite{Fleischhauer05RMP,Novikova12review}. Then, a \emph{``write''} process is applied, which maps the envelope of a probe pulse onto a spin wave of spatially distributed atomic coherences $\rho_{12}(z,0)$ in the medium, which contains all information about the probe pulse \cite{Novikova12review}
\begin{equation}
\mathcal{E}_{\text{probe}}(z,0)\to\sqrt{N/V}\rho_{12}(z,0)\e^{\i\Delta k z},
\end{equation}
where $\mathcal{E}_{\text{probe}}(z,t)$ is the electric field envelope of the probe pulse, $N/V$ is the number density of atoms, $\rho_{12}(z,t)$ is the coherence at position $z$ in the ensemble, $\Delta k$ the phase mismatch between to probe and control beams, e.g., due to the geometry of the experiment (see Fig. \ref{Fig:fig4}). Thus, the phase of $\xi_{12}(z,t)$ can be spatially retarded and vary along $z$ after the \emph{``write''} process.

%

\begin{table*}[t]
\centering
\caption{EIT light storage rephasing efficiency $\eta_{\text{ls}}$ for rephasing sequences with equal pulse separation \cite{RDD_review12Suter,Genov2014PRL,Genov17PRL}.
\vspace{0.3cm}
} 
\begin{tabular}{l l l l} 
\hline\hline 
Pulses~ & Sequence~ & Phases~~ & Rephasing efficiency $\eta_{\text{ls}}$~~~~~~~~~~~~~~~\\ 
\hline 
2 & CPMG & $(0, \phi_2)$ & $\langle 1 - 2\epsilon+\epsilon^2 \rangle^2$\\
4 & XY4 & $(0, 1,0,1)\pi/2$ & $\langle 1 - 4\epsilon^2 +4\epsilon^3-\epsilon^4 \rangle^2$\\
4 & UR4 & $(0, 1,1,0)\pi$ & $\langle 1 - 4\epsilon^2 +4\epsilon^3-\epsilon^4 \rangle^2$\\
6 & UR6 & $(0, 2,0,0, 2,0)\pi/3$ &  $\langle 1 - 2\epsilon^3 -4\epsilon^4+8\epsilon^5-3\epsilon^6 \rangle^2$\\
8 & XY8 & $(0, 1,0,1,1,0,1,0)\pi/2$ & $\langle 1 - 8\epsilon^3 +4\epsilon^4+48\epsilon^6-80\epsilon^7+35\epsilon^8 \rangle^2$\\
8 & UR8 & $(0,1,3,2,2,3,1,0)\pi/2$ &  $\langle 1 - 4\epsilon^4 +8\epsilon^5-16\epsilon^6+16\epsilon^7-5\epsilon^8 \rangle^2$\\
10 & [U5a]$^2$ & $(0, 5, 2,5,0,0, 5, 2,5,0)\pi/6$ & $\langle 1 - 2(11-6\sqrt{3}) \epsilon ^3+4(11-6\sqrt{3}) \epsilon ^4-12(2-\sqrt{3})\epsilon^5+\text{O}(\epsilon^6)\rangle^2$\\
10 & [KDD]$^2$ & $(1, 0, 3,0,1,1, 0, 3,0,1)\pi/6$ & $\langle 1 - 2(11+6\sqrt{3}) \epsilon ^3+4(11+6\sqrt{3}) \epsilon ^4-12(2+\sqrt{3})\epsilon^5+\text{O}(\epsilon^6)\rangle^2$\\
10 & UR10 & $(0,4,2,4,0,0,4,2,4,0)\pi/5$~~~& $\langle 1 - 2\epsilon^5 -2(3-\sqrt{5})\epsilon^6+8(2-\sqrt{5})\epsilon^7-2(11-5\sqrt{5})\epsilon^8+4(5-\sqrt{5})\epsilon^9-7\epsilon^{10} \rangle^2$\\
\hline 
\end{tabular}
\label{Table:sequences} 
\end{table*}

During a storage time $T_{\text{st}}$, we \emph{``rephase''} the coherences by applying a sequence of time-separated pulses to counter the effect of dephasing. The density matrix at a time $T_{\text{st}}$ along $z$ is $\rho(z,T_{\text{st}})$, given by Eq. \eqref{densityM_rephasing}.

The spin wave is then read out by applying a control read pulse to beat with the atomic coherences and generate a signal pulse on the $|1\rangle\leftrightarrow|e\rangle$ transition. This is usually termed the \emph{``read''} process, which in the perfect case for forward readout maps back the spin wave onto an electromagnetic field on the probe  pulse transition \cite{Fleischhauer05RMP}
\begin{equation}
-\sqrt{N/V}\langle\rho_{12}(z,T_{\text{st}})\rangle\e^{\i\Delta k z}\to \mathcal{E}_{\text{signal}}(z,T_{\text{st}}).
\end{equation}
If $\langle\rho_{12}(z,T_{\text{st}})\rangle=\rho_{12}(z,0)$, e.g., with perfect rephasing pulses, and the control read pulse is in the same direction as the control write pulse, the retrieved pulse will be a perfect time-inverted copy of the stored pulse, propagating in the same direction due to the phase-matching condition \cite{Surmacz08PRA,Novikova12review}.
However, the rephasing pulses are usually not perfect.
In the following we derive a simplified expression for the rephasing efficiency for EIT light storage, which we then use to analyze the effect of pulse imperfections in the main text.

First, we assume without loss of generality that the probe pulse envelope $\mathcal{E}_{\text{probe}}(z,t)$ is real, so the coherence phase after the \emph{``write''} process is $\xi(z,0)=-\Delta k z$. Perfect phase matching for forward readout 
requires
\begin{equation}
\langle\rho_{12}(z,T_{\text{st}})\rangle\e^{\i\Delta k z}=|\langle\rho_{12}(z,T_{\text{st}})\rangle|
\end{equation}
that $\xi(z,T_{\text{st}})=-\Delta k z=\xi(z,0)$. Microscopically, phase matching implies that the individual atomic dipoles along the axis of propagation are properly phased, so that their emission adds up coherently in the propagation direction of the signal field \cite{Surmacz08PRA}. If we assume perfect EIT conditions after readout to avoid reabsorption, e.g., two-photon resonance and negligible population in states $|e\rangle$ and $|2\rangle$, it is known that
the signal pulse envelope at distance $L$ (end of our storage medium) yields \cite{Boyd2008NonlinOpt}:
\begin{equation}\label{Eq:Esignal}
\mathcal{E}_{\text{signal}}(L,T_{\text{st}})\sim\int_{0}^{L}\langle\rho_{12}(z,T_{\text{st}})\rangle\e^{\i\Delta k z}\d z.
\end{equation}
The light storage efficiency is defined by the ratio of the energy of retrieved photons after storage vs. the energy of the photons of the input probe pulse \cite{Fleischhauer05RMP,Novikova12review}
\begin{equation}
\eta_{\text{ls}}=\frac{\int_{T_{\text{st}}}^{\infty}| \mathcal{E}_{\text{signal}}(z=L,t)|^2\d t}{\int_{-\infty}^{0}| \mathcal{E}_{\text{probe}}(z=0,t)|^2\d t},
\end{equation}
where $z=0$ and $z=L$ are the beginning and end of the storage medium.
We assume assume that EIT light storage \emph{``write''} and \emph{``read''} procedures are perfect, and there is no change in the duration of the probe field after the storage. Then, the storage efficiency can be approximated by the ratio of magnitudes of the time-averaged Poynting vectors of the probe field after and before the storage \cite{Boyd2008NonlinOpt}. We also assume that the magnitude of the Poynting vector is proportional to $|\mathcal{E}_{\text{signal}}(L,t)|^2$, which should be feasible, e.g., for a constant probe pulse envelope.
Then, we use Eq. \eqref{Eq:Esignal} to obtain
\begin{equation}\label{LS_efficiency_all_appendix}
\eta_{\text{ls}}=\frac{|\int_{0}^{L}\langle\rho_{12}(z,T_{\text{st}})\rangle\e^{\i\Delta k z} \d z|^2}{|\int_{0}^{L}\rho_{12}(z,0)\e^{\i\Delta k z}\d z|^2}.
\end{equation}
We note that this expression does not take into account reabsorption of the signal field, e.g., in case of imperfect EIT because of significant population in state $|2\rangle$ after the rephasing sequence. Nevertheless, it provides a good qualitative estimate for the relative change in the storage efficiency for sequences of pulses when we change the phases of the latter. Next, we assume for simplicity that $|\rho_{12}(z,0)|$ is constant along $z$ (or does not vary much along $z$ for a length for $\Delta k z$ changes from $0$ to $2\pi$), which allow us to simplify Eq. \eqref{LS_efficiency_all_appendix} to obtain
\begin{equation}
\eta_{\text{ls}}\approx\left|\frac{1}{L}\int_{0}^{L}\frac{\langle\rho_{12}(z,T_{\text{st}})\rangle}{|\rho_{12}(z,0)|}\e^{\i\Delta k z} \d z\right|^2.
\end{equation}
It is obvious that $\langle\rho_{12}(z,T_{\text{st}})\rangle=\rho_{12}(z,0)=|\rho_{12}(z,0)|\exp{(-\i\Delta k z)}$ implies that $\eta_{\text{ls}}=1$.

Alternatively, we can further simplify the calculation of the integrals in Eq. \eqref{LS_efficiency_all_appendix} by change of variables $\Delta k z\to -\xi(z,0)$ if we assume that the phases of the coherence after the \emph{``write''} process $\xi(z,0)=-\Delta k z$ are equally distributed between $0$ and $2\pi$, so
\begin{subequations}
\begin{align}
\eta_{\text{ls}}&=\frac{|\int_{0}^{2\pi}\langle\rho_{12}(z,T_{\text{st}})\rangle\e^{-\i\xi(z,0)} \d \xi(z,0)|^2}{|\int_{0}^{2\pi}\rho_{12}(z,0)\e^{-\i\xi(z,0)}\d \xi(z,0)|^2}\\
&\approx\frac{|\int_{0}^{2\pi}\langle\rho_{12}(z,T_{\text{st}})\rangle\e^{-\i\xi(z,0)} \d \xi(z,0)|^2}{|2\pi\rho_{12}(z,0)|^2}\\
&=\left|\frac{1}{2\pi}\int_{0}^{2\pi}\frac{\langle\rho_{12}(z,T_{\text{st}})\rangle}{\rho_{12}(z,0)} \d \xi(z,0)\right|^2,
\end{align}
\end{subequations}
where we used that $|\rho_{12}(z,0)|=\rho_{12}(z,0)\e^{-\i\xi(z,0)}$ is approximately constant for the small distance along $z$ where the initial coherence phase $\xi(z,0)=-\Delta k z$ changes from $0$ to $2\pi$. We note that the above expression for the rephasing efficiency is also valid whenever averaging over a phase distribution of initial coherences is required, i.e., also beyond the specific case of spatial retardation.

%
\subsection{Rephasing efficiency in EIT light storage for some robust sequences}
\label{Appendix:Efficiency_robust_sequences}

We use the theoretical approach from Sec. \ref{Subsec:Theory_LS_Efficiency} and provide explicit analytical formulas for the rephasing efficiency of several robust rephasing sequences in Table \ref{Table:sequences}.




\begin{thebibliography}{99}


\bibitem{Lvovsky09NPhot} A.~I.~Lvovsky, B.~C.~Sanders, and W.~Tittel, Nat. Phot. \textbf{3}, 706-714 (2009).

\bibitem{Fleischhauer05RMP} M.~Fleischhauer, A.~Imamoglu, and J.~P.~Marangos, Rev. Mod. Phys. \textbf{77}, 633 (2005).

\bibitem{MarangosHalfmann09Optics} J.~P.~Marangos and T.~Halfmann, \emph{Electromagnetically Induced Transparency}, in Handbook of Optics Vol. IV (McGraw-Hill Professional, New York, 2009).

\bibitem{Heinze13PRL} G.~Heinze, C.~Hubrich, T.~Halfmann, Phys. Rev. Lett. \textbf{111}, 033601 (2013).

\bibitem{Lovric13PRL} M.~Lovric, D.~Suter, A.~Ferrier, and P. Goldner, Phys. Rev. Lett. \textbf{111}, 020503 (2013).

\bibitem{Zhong15Nature} M.~Zhong, M.~P.~Hedges, R.~L.~Ahlefeldt, J.~G.~Bartholomew, S.~E.~Beavan, S.~M.~Wittig, J.~J.~Longdell, and M.~J.~Sellars, Nature (London) \textbf{517}, 177 (2015).

\bibitem{Jobez15PRL} P.~Jobez, C.~Laplane, N.~Timoney, N.~Gisin, A.~Ferrier, P.~Goldner, and M.~Afzelius, Phys. Rev. Lett. \textbf{114}, 230502 (2015).

\bibitem{Mieth16PRA} S.~Mieth, G.~T.~Genov, L.~P.~Yatsenko, N.V.~Vitanov, and T.~Halfmann, Phys. Rev. A \textbf{93}, 012312 (2016).

\bibitem{Schraft16PRL} D.~Schraft, M.~Hain, N.~Lorenz, and T.~Halfmann, Phys. Rev. Lett. \textbf{116}, 073602 (2016).

\bibitem{DegenRMP2017} C.~L.~Degen, F.~Reinhard, and P.~Cappellaro, Rev. Mod. Phys. \textbf{89}, 035002 (2017).

\bibitem{HaeffnerPR2008} H.~H{\"a}ffner, C.~F.~Roos, and R.~Blatt, Phys. Reports \textbf{469}, 155-203 (2008).

\bibitem{PiltzScience2016} Ch.~Piltz, Th.~Sriarunothai, S.~S.~Ivanov, S.~Wolk and Ch.~Wunderlich, Science Advances \textbf{2}, e1600093 (2016).


\bibitem{RDD_review12Suter} A.~Souza, G.~A.~Alvarez, and D.~Suter, Phil. Trans. R. Soc. A \textbf{370}, 4748-4769 (2012), and references therein.

\bibitem{Torosov11PRA} B.~T.~Torosov and N.~V.~Vitanov, Phys. Rev. A \textbf{83}, 053420 (2011).

\bibitem{Torosov11PRL} B.~T.~Torosov, S.~Gu\'erin and N.~V.~Vitanov, Phys. Rev. Lett. \textbf{106}, 233001 (2011).

\bibitem{Schraft13PRA} D.~Schraft, T.~Halfmann, G.~T.~Genov, and N.V.~Vitanov, Phys. Rev. A \textbf{88}, 063406 (2013).

\bibitem{Casanova15PRA} J.~Casanova, Z.-Y.~Wang, J.~F.~Haase, and M.~B.~Plenio, Phys. Rev. A \textbf{92}, 042304 (2015).

\bibitem{Van-Damme17PRA} L.~Van-Damme, D.~Schraft, G.~T.~Genov, D.~ Sugny, T.~Halfmann, and S.~Gu\'erin, Phys. Rev. A \textbf{96}, 022309 (2017).

\bibitem{Genov2014PRL} G.~T.~Genov, D.~Schraft, T.~Halfmann, N.~V.~Vitanov, Phys. Rev. Lett. \textbf{113}, 043001 (2014).

\bibitem{Genov17PRL} G.~T.~Genov, D.~Schraft, N.V.~Vitanov, and T.~Halfmann, Phys. Rev. Lett. \textbf{118}, 133202 (2017).

\bibitem{CPMG_papers} H.~Y.~Carr and E.~M.~Purcell, Phys. Rev. \textbf{94}, 630 (1954);
S.~Meiboom and D.~Gill, Rev. Sci. Instrum. \textbf{29}, 688-691 (1958).

\bibitem{Levitt84NMR} M.~H.~Levitt, Prog. NMR Spectrosc. \textbf{18}, 61 (1986);
R.~Freeman, \emph{Spin Choreography} (Spektrum, Oxford, 1997).

\bibitem{Cruzeiro16JMO} E.~Z.~Cruzeiro, F.~Fr{\"o}wis, N.~Timoney and M.~Afzelius, J. Mod. Opt. \textbf{63}, 2101-2113 (2016).

\bibitem{Hahn50PR} E.~L.~Hahn, Phys. Rev. \textbf{80}, 580 (1950).

\bibitem{NMR_literature} A.~Abragam, \emph{The Principles of Nuclear Magnetism} (Oxford
University Press, Oxford, 1961);
C.~P.~Slichter, \emph{Principles of Magnetic Resonance} (Springer, Berlin, 1990).

\bibitem{Wong83PRB} N.~C.~Wong, E.~S.~Kintzer, J.~Mlynek, R.~G.~DeVoe, and R.~G.~Brewer, Phys. Rev. B \textbf{28}, 4993 (1983).

\bibitem{Lauro11PRA} R.~Lauro, T.~Chaneli\'ere, and J.-L.~Le Gou{\"e}t, Phys. Rev. B \textbf{83}, 035124 (2011).

\bibitem{Mieth12PRA} S.~Mieth, D.~Schraft, T.~Halfmann, and L.~P.~Yatsenko, Phys. Rev. A \textbf{86}, 063404 (2012).

\bibitem{Pascual-Winter13NJP} M.~F.~Pascual-Winter, R.-C.~Tongning, T.~Chaneli\'ere, and J.-L.~Le Gou{\"e}t, New J. Phys. \textbf{15}, 055024 (2013).

\bibitem{Shore_literature} B.~W.~Shore, \emph{The Theory of Coherent Atomic Excitation} (Wiley, New York, 1990);  B.~W.~Shore, \emph{Manipulating Quantum Structures Using Laser Pulses} (Cambridge University Press, Cambridge, 2014).


\bibitem{Novikova12review} I.~Novikova, R.~L.~Walsworth, and Y.~Xiao, Laser Phot. Rev. \textbf{6}, 333 (2012).

\bibitem{Heinze14PRA} G.~Heinze, C.~Hubrich, T.~Halfmann, Phys. Rev. A \textbf{89}, 053825 (2014).

\bibitem{Mieth14OE} S.~Mieth, A.~Henderson, and T.~Halfmann, Opt. Express, \textbf{22}, 11182 (2014).

\bibitem{Serrano2018NattComm} D.~Serrano, J.~Karlsson, A.~Fossati, A.~Ferrier and P.~Goldner, Nat. Comm. \textbf{9}, 2127 (2018).

\bibitem{Surmacz08PRA} K.~Surmacz, J.~Nunn, K.~Reim, K.~C.~Lee, V.~O.~Lorenz, B.~Sussman, I.~A.~Walmsley, and D.~Jaksch, Phys. Rev. A \textbf{78}, 033806 (2008).





\bibitem{PezzeRMP2018} L.~Pezz\'e, A.~Smerzi, M.~K.~Oberthaler, R.~Schmied, and P.~Treutlein, Rev. Mod. Phys. \textbf{90}, 035005 (2018).

\bibitem{Arecchi72PRA} F.~T.~Arecchi, E.~Courtens, R.~Gilmore, and H.~Thomas, Phys. Rev. A \textbf{6}, 2211 (1972).

\bibitem{Friedberg07LPL} R.~Friedberg and J.~T.~Manassah, Laser Phys. Lett. \textbf{4}, 900911 (2007).



\bibitem{Demkov-Kunike} Yu.~N.~Demkov, M.~Kunike, Vestn. Leningr. Univ. Fiz. Khim. \textbf{16}, 39 (1969);
F.~T.~Hioe, C.~E.~Carroll, Phys. Rev. A \textbf{32}, 1541 (1985);
J.~Zakrzewski, Phys. Rev. A \textbf{32}, 3748 (1985);
K.-A. Suominen, B.M. Garraway, Phys. Rev. A \textbf{45}, 374 (1992).

\bibitem{GenovOptComm2011} G.~T.~Genov, A.~A.~Rangelov, and N.~V.~Vitanov, Opt. Comm. \textbf{284}, 2642 (2011).

\bibitem{Boyd2008NonlinOpt} R.~W.~Boyd, \emph{Nonlinear Optics}, 3rd ed. (Academic Press, Burlington, MA, 2008).

\end{thebibliography}
\end{document}